\documentclass[iop,apj]{emulateapj}

\usepackage{amssymb,amsmath,graphicx,epstopdf,url,nicefrac}
\usepackage{hyperref}

\usepackage{natbib}
\slugcomment{Submitted to ApJ}

\newcommand{\NH}{\mbox{${\rm N}_{\rm H}$}} 

\newcommand{\msun}{${\rm M_{\odot}}$}
\newcommand{\msunyr}{${\rm M_{\odot}}~\rm{yr}^{-1}$}
\newcommand{\Lxiso}{$L_{\rm X,iso}$}

\renewcommand{\apj}{{ApJ}} 	    
\renewcommand{\apjl}{{ApJL}}  
\renewcommand{\aap}{{A\&A}}   
\renewcommand{\mnras}{{MNRAS}} 	            
\renewcommand{\nat}{{Natur}}               
\renewcommand{\pasj}{{PASJ}}
\renewcommand{\ssr}{{SSRv}}   
 
\newcommand{\chandra}{\mbox{\it Chandra}}	   
\newcommand{\swift}{{\it Swift}}

\def\simlt{\mathrel{\hbox{\rlap{\hbox{\lower4pt\hbox{$\sim$}}}\hbox{$<$}}}}
\def\simgt{\mathrel{\hbox{\rlap{\hbox{\lower4pt\hbox{$\sim$}}}\hbox{$>$}}}}

\newcommand{\transient}{{Swift\,J1644+57}}

\shorttitle{X-ray Light Curve of Swift~J1644+57}
\shortauthors{Mangano et al.}

\begin{document}

\title{The Definitive X-ray Light Curve of Swift J164449.3+573451}

\author{V.\ Mangano\altaffilmark{1},
D. N. Burrows\altaffilmark{1},
B.\ Sbarufatti\altaffilmark{2,1},
J. K. Cannizzo\altaffilmark{3,4}}

\altaffiltext{1}{Department of Astronomy \& Astrophysics, The Pennsylvania State
  University, 525 Davey Lab, University Park, PA 16802, USA; \textit{vm.himawari@gmail.com}}

\altaffiltext{2}{INAF -- Osservatorio Astronomico di Brera, Via Bianchi 46,
    23807 Merate, Italy}

\altaffiltext{3}{CRESST and Astroparticle Physics Laboratory NASA/GSFC, Greenbelt, MD 20771, USA}

\altaffiltext{4}{Department of Physics, University of Maryland, Baltimore County, Baltimore, MD 21250, USA}

\begin{abstract}
On 2011 March 28, the \swift\ Burst Alert Telescope  
triggered on an object that had no analog in over 
six years of \swift\ operations. 
Follow-up observations by the \swift\ X-ray Telescope (XRT)
found a new, bright X-ray source covering three orders of magnitude 
in flux over the first few days, that was much more persistent 
(and variable) than gamma-ray burst afterglows. 
Ground-based spectroscopy found a redshift of 0.35, 
implying extremely high luminosity, with integrated 
isotropic-equivalent energy output in the X-ray band 
alone exceeding $10^{53}$ ergs in the first two weeks 
after discovery. Strong evidence for a collimated outflow 
or beamed emission was found.
The observational properties of this object are unlike 
anything ever before observed.
We interpret these unique properties as the result of 
emission from a relativistic jet produced in the aftermath 
of the tidal disruption of a main sequence star by a massive 
black hole (BH) in the center of the host galaxy. 
The source decayed slowly as the stellar remnants were 
accreted onto the BH, before abruptly shutting off. 
Here we present the definitive XRT team light curve for 
Swift~J164449.3+573451 and discuss its implications.  
We show that the unabsorbed flux decayed roughly as a $t^{-1.5}$  
power law up to 2012 August 17.
The steep turnoff of an order of magnitude in 24 hr 
seems to be consistent with the shutdown of the jet 
as the accretion disk transitioned from a thick disk to a thin disk.  
\swift\ continues to monitor this source in case the jet reactivates.
\end{abstract}

\keywords{accretion, accretion discs - black hole physics - 
galaxies: active - galaxies: jets - 
X-rays: galaxies - X-rays: individual (Swift J164449.3+573451)} 

\defcitealias{Burrows11_SwiftJ1644}{Paper~I}

\section{Introduction}

On 2011 March 28, the \swift\ satellite \citep{Gehrels04_Swift} 
triggered on a new source, \break Swift~J164449.3+573451 
(\transient\ hereafter).   
\transient\ triggered the \swift\ Burst Alert Telescope 
\citep[BAT;][]{Barthelmy05} 
three more times over the next two days, 
a clear indication that it was not a gamma-ray burst (GRB).  
Observations by the \swift\ X-ray Telescope 
\citep[XRT;][]{Burrows05_XRT} 
revealed an extremely variable, but previously undetected X-ray source 
\citep{Burrows11_SwiftJ1644}, which was found to be located at the center 
of a galaxy at a redshift of 0.354 \citep{Levan11_SwiftJ1644}.  
A consensus quickly developed that this object was a highly beamed 
tidal disruption event 
\citep[TDE;][]{Bloom11_SwiftJ1644, Levan11_SwiftJ1644, 
Burrows11_SwiftJ1644, Zauderer11_SwiftJ1644}.

Because the extinction in the host galaxy is so high, the value 
of extinction adopted drives the inferred slope of the inherent 
optical-NIR portion of the spectrum, which in turn drives the 
interpretation of the X-ray spectrum.  
\citet[][hereafter \citetalias{Burrows11_SwiftJ1644}]{Burrows11_SwiftJ1644} 
found that the host extinction was $A_V \approx 4.5 $, which led 
to a positive slope of the intrinsic spectrum in the optical band 
and our conclusion that the optical and X-ray bands were both part 
of the same synchrotron component.  
This interpretation required that the radio emission originate 
in a different emission region, which we proposed was in the external shock.  
 \citet{Zauderer11_SwiftJ1644} and \citet{Metzger12_SwiftJ1644} came 
to similar conclusions, based on their radio observations.  
 \citet{Bloom11_SwiftJ1644}, on the other hand, concluded 
that the radio and NIR/optical emission originated in the 
same synchrotron component. 
 
Prior to the discovery of \transient, relativistic jets 
created by tidal disruption around dormant black holes (BHs)
had only rarely been discussed  in the literature on TDE theory
(e.g., \citealt{Giannios11}; cf. \citealt{Rees88,Ayal00,Bogdanovic04,Gomboc05,
Brassart08,Strubbe09,Guillochon09,Brassart10}). 
These earlier works generally considered scenarios in which a solar mass star 
is tidally disrupted by a supermassive black hole (SMBH), 
roughly half of the stellar mass is captured with 
the remainder of the disrupted star ejected into hyperbolic orbits, 
and visible, UV, and X-ray flares are produced by the impulsive 
compression of the star, the resulting accretion disk, 
and/or the outflowing material.
In retrospect, it is not surprising that a jet could be formed 
in a TDE, as the SMBHs in an active galactic nucleus (AGN) produce jets 
under conditions of ongoing accretion.  
The unique aspect of this discovery was the detailed observation 
in the X-ray band of the turn-on of a jet in a previously dormant SMBH.

The discovery of this event produced a flurry of publications 
providing interpretation of the data.  
While most authors (including \citetalias{Burrows11_SwiftJ1644}) 
argued for a solar mass star being tidally disrupted by a SMBH, 
\citet{Krolik11_SwiftJ1644} preferred a white dwarf being disrupted 
by an intermediate mass BH in order to explain the rapid 
time variability.  
\citet{Cannizzo11_SwiftJ1644} suggested that the star was
on an orbit with a very small impact parameter, 
and the resulting plunge deep into the SMBH potential well 
totally disrupted the star, which was then completely swallowed
by the BH.  


The purpose of the present work is the complete analysis of
the \swift/XRT  follow-up data of \transient\ from the beginning to
the end of the outburst, which occurred on 2012 August 17th, 507 days
after the initial BAT trigger, and the following monitoring
observations by \swift/XRT and \chandra/ACIS. 
We present the updated complete version 
of the flux light curve of the event and its analysis
and interpretation.


This paper is organized as follows.
In Section \ref{sec:obs} we describe the data set.
In Section \ref{sec:redu} we describe in detail
the data reduction and analysis procedures performed 
to obtain the final X-ray light curve of \transient.
In Sections \ref{sec:lcfit} and \ref{sec:dips} we analyze 
the light curve structure and late time behavior. 
In Section \ref{sec:discussion} we discuss our results 
and their implications in the context of tidal disruption models 
of a star by a SMBH associated to relativistic jet ejection.
Finally, in Section \ref{sec:conclusions} we summarize our findings 
and conclusions. 

Throughout this paper 
the errors on count rates are at 1$\sigma$ 
\citep[e.g. in light curves;][]{Evans07,Evans09}, and
the quoted uncertainties on model parameters
are given at 90\%  confidence level for one interesting parameter 
(i.e., $\Delta \chi^2 =2.71$) unless otherwise stated.
Times $t$ are referenced to the initial BAT trigger $T_0$
as $t=T-T_0$, unless otherwise specified. 
We adopt a standard cosmology model with $H_0 = 70$ km s$^{-1}$ Mpc$^{-1}$, 
$\Omega_{\rm M} = 0.3$, $\Omega_\Lambda = 0.7$.

\section{Observations}
\label{sec:obs}

\transient\ triggered the \swift/BAT twice on
2011 March 28th, resulting in automated observations by \swift/XRT.
It was intensively observed by \swift\ for the first four months 
and then was regularly monitored with single snapshots 
of 1 ks exposure per day,
at diminishing rate per week.
The whole \swift\ follow-up and monitoring campaign of 
\transient\ from discovery up to 
three years later 
consists of $\sim$700 sequences and four different trigger numbers:
450158 (seq. 0$-$7, from 2011 March 28 to 2011 March 30),
31955 (seq. 2-255, from 2011 March 31 to 2011 December 06),
32200 (seq. 1-237, from 2011 December 07 to 2012 August 15), and
32526 (seq. 1-200, from 2012 August 16 to 2014 March 28).
The total XRT exposure in the data set is 
2 Ms in Photon Counting (PC) mode and  
320 ks in Windowed Timing (WT) mode 
\citep[see][for a description of the XRT instrument modes]{Burrows05_XRT}.

We also re-analyze the 24.7 ks \chandra-ACIS ToO observation 
(PI: Tanvir),  performed on 2012 November 26, and present analysis 
results for a second ACIS observation of 27.7 ks exposure  performed 
by \chandra\ on 2015 February 17  (PI: Levan).

The reference time used to plot all the light curves 
and parameter evolution curves
is the time  of the first BAT trigger $T_0$, corresponding to
$T_0$ $=$ 2011 March 28 12:57:45.201 UTC $=$ 55648.5401 MJD.

\section{Data Reduction and Analysis}
\label{sec:redu}

The XRT data have been reprocessed with the HEASOFT 6.15.1 package,
and the latest calibration files (version 20140610) have been used 
for pile-up correction and spectral response matrices production.

For light curve and hardness ratio curve extraction 
(see Sections \ref{subsec:rate_lc}  and \ref{subsec:hr_lc}), 
an annular region centered at the source position
with variable inner radius has been used 
when pile-up correction was required, 
and a circular region with outer radius
decreasing with the average brightness 
,of the source has been used in all other cases.
A minimum extraction radius of 10 pixels (23.6\arcsec)
has been used starting from 2012 August 17 
(day 508 since the trigger, sequence 00032526002). 
The background region for observations in PC mode 
has been selected as a set of circular
regions surrounding the source where no serendipitous
source was detected in the image of the summed PC
observations (2 Ms exposure). 
The background region used for observations 
in WT mode in \citetalias{Burrows11_SwiftJ1644} 
was an annular region with 65\arcsec\
inner radius and 70\arcsec\ outer radius. 
Here we preferred a constant background subtraction,
with background rate estimated from average background
spectra described below. This procedure allows 
for more reliable results, especially in the soft (highly absorbed)
band 0.3$-$10 keV.

To extract time resolved spectra 
(see Section \ref{subsec:spectra}) 
and average spectra in or outside the light curve dips 
(see Section \ref{sec:dips}),
source regions matching the ones used for the 
light curve in each corresponding time interval 
have been generally used. 
PC background spectra have been extracted 
from simultaneous data 
using an annular region centered on the source position 
with inner radius of 119 pixels ($\sim$281\arcsec) 
and outer radius of 153 pixels ($\sim$362\arcsec), 
unless otherwise stated.
For the WT spectra, 
in order to maximize the background statistics,
only two background spectra have been created:
{\it i)} an average background spectrum (15 ks exposure)
from the summed WT observations in sequences 00450158000-06, 
used for the early ($<$10 d) WT spectra;
{\it ii)} an average background spectrum (296.5 ks exposure)
from the summed WT observations in sequences 00031955014-40,
used in all other cases\footnote{Note that no WT observations 
corresponding to trigger numbers 32200 or 32526 are present 
in our data set.}. Both background spectra have been 
extracted from an annular region centered on the source 
with 85 pixels ($\sim$200\arcsec) inner radius 
and 118 pixels ($\sim$279\arcsec) outer radius.
Spectra have been generally binned to at least 20 counts per energy bin
to allow for fitting in $\chi^2$ statistics within {\tt xspec}, unless
otherwise stated.
Ancillary response files were generated with the task 
{\tt xrtmkarf} within HEASOFT, and account for different 
extraction regions and point-spread function corrections. 

\chandra-ACIS data have been analyzed
with  the  CIAO  software  package (v4.7),  
using  the  calibration  database  CALDB  (v4.6.7)  
and standard ACIS data filtering. 
We used the {\tt wavedetect} task for source detection.
The source and background regions used for count rate estimation and 
spectral extraction are a circle centered on the source position with 1\farcs5 radius,
and a source free annular region with 15\arcsec\ inner radius and 45\arcsec\ outer radius, 
respectively.
For spectral analysis the data were binned to 1 count per energy 
bin to allow  for fitting with Cash statistics \citep{Cash79} in {\tt xspec}
(with $\chi^2$ statistic test evaluation enabled).
\chandra\ results are presented and discussed in Section \ref{subsec:residual}.

\subsection{Count Rate Light Curve}
\label{subsec:rate_lc}


%
\begin{figure}[ht]
\resizebox{\hsize}{!}{\includegraphics[angle=0]{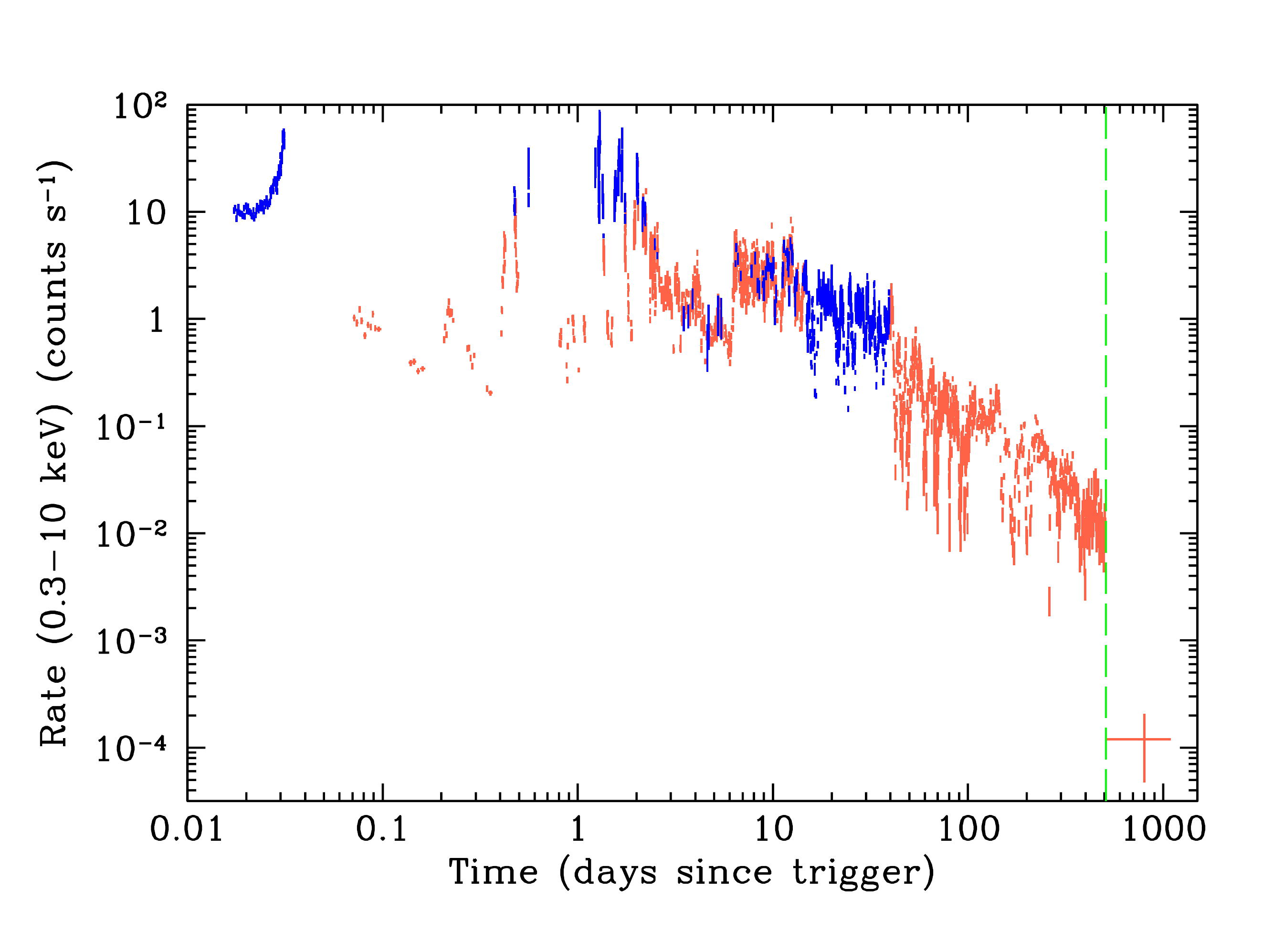} }
\caption[]{X-ray light curve of \transient\ in the 0.3$-$10 keV band.  
Blue points are data taken in WT mode; red points are data in PC mode.
Details of the early X-ray light curve are presented 
and discussed in \citetalias{Burrows11_SwiftJ1644}.  
Here we present the entire light curve measured by the \swift/XRT.  
Note the strong variability during the first three days, 
the decay beginning around day 7 and punctuated by deep dips 
at irregular intervals, and the abrupt decline on day 508 
marked by the vertical dashed green line. 
See Figure~\ref{fig:Aug-Sept} for details.
The error of the late detection after the dashed green line 
is at the 99\% confidence level.
\label{fig:XRT_lc}}
\end{figure}


%
\begin{figure}[ht]
\resizebox{\hsize}{!}{\includegraphics[angle=0]{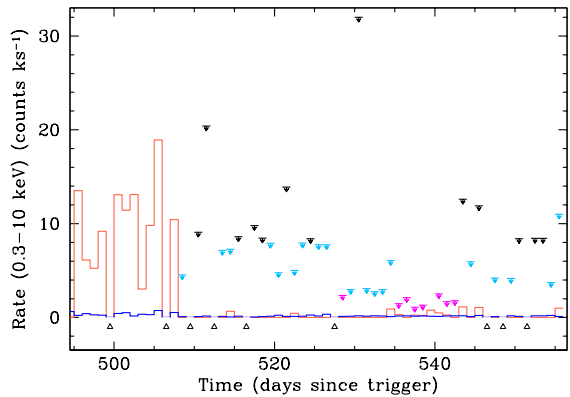} }
\caption[]{X-ray light curve of \transient\ in the 0.3$-$10 keV band 
during 2012 August and September, binned in days.  
The red histogram represents the count rate measured in the source region.
The blue histogram represents the background rate rescaled to 
the source region area from the rate measured in the background region.
Errors are not plotted for a better readability of the figure.
The line of black upward triangles at the bottom of the plot 
indicates days in which no observation  was performed.
The segments marked with downward arrows are the 3$\sigma$ upper limits 
on source count rate when the source was not detected 
with the {\tt ximage} command {\tt detect},
obtained by running the command {\tt sosta} 
in the position of \transient\ with user defined 
source and background regions. 
The black arrows correspond to observations with less than 1~ks exposure, 
the light blue arrows to 1-3~ks exposure, 
and the magenta arrows to more than 3~ks exposure.
The plot shows that the count rate drops abruptly on day 508 (2012 August 17) and stays below
the 3$\sigma$ upper limit later on. 
\label{fig:Aug-Sept}}
\end{figure}


The 0.3$-$10 keV count rate light curve shown in 
Figure~\ref{fig:XRT_lc} was produced by binning the 
net source counts in order to have at least 200 
counts per time bin at early times,
and progressively decreasing the level of 
required counts per bin
down to at least 10 counts per bin at late times,
up to sequence 00032526001, observed on 2012 August 16,  
507 days after the trigger.
On day 508 the source emission suddenly dropped below 
the detection threshold for a 1$-$2 ks single snapshot 
observation \citep{Sbarufatti12_ATel_SwiftJ1644}. 
A detailed plot of the abrupt drop is shown in 
Figure~\ref{fig:Aug-Sept},
where data from 2012 August and September have been binned 
daily.
The daily count rate measured in the source region 
(red histogram)
is compared to the expected background count rate 
in the source region on the same day
(blue histogram).  When \transient\ is not detected, 
the 3$\sigma$ upper limit on source count rate is shown. 
The plot shows that the count rate drops abruptly on day 508 
(2012 August 17).
By comparing the count rate on day 507 and the level of the
deepest upper limit we see in Figure~\ref{fig:Aug-Sept}
(on day 537) we can state that we had a drop of more than
a factor 11 in one month, i.e. a fraction of 0.06 of the total
time elapsed.

The very last point in the light curve of Figure~\ref{fig:XRT_lc}
is the detection obtained from the summed 191
monitoring observations performed from 2012 August 17th 
(marked by the green dashed vertical line) 
to 2014 March 28th, for a total of 284.5 ks exposure.
We extracted the source counts from a circular region
of 10 pixels radius (314 pixel$^2$ area) and estimated 
the background contribution from counts found in the
background region (21,143 pixel$^2$ area, specifically selected to avoid 
contamination from any field source detected in the whole PC data set)
and rescaled to the source region area.
We obtained 78 total counts in the source region in the 0.3$-$10 keV band
with an estimated background of 44 counts 
and applied the technique of \citet{Kraft91} 
to calculate the upper and lower 99\% confidence 
limits on source counts. 
These correspond to an average count rate of 
$1.2^{+0.9}_{-0.7} \times 10^{-4}$~s$^{-1}$.
The source is also detected on the summed image 
of the post-drop monitoring 
with the {\tt ximage} command {\tt sosta}, centered 
on the position of source with user defined source 
and background regions, at the level 
$\rm{(}1.5 \pm 0.4\rm{)} \times 10^{-4}$~s$^{-1}$
with 3.8$\sigma$ significance.

\subsection{Hardness Ratio}
\label{subsec:hr_lc}


%
\begin{figure}[b]
\resizebox{\hsize}{!}{\includegraphics[angle=0]{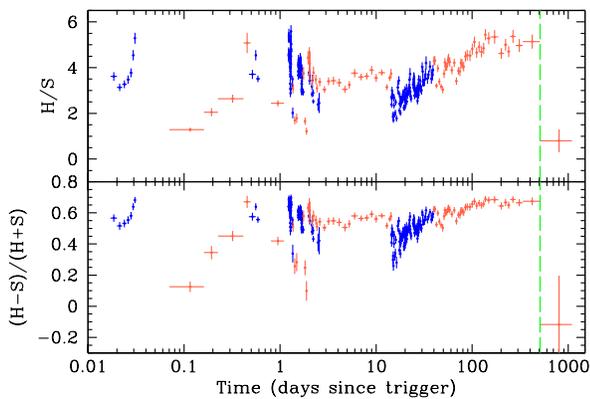} }
\caption[]{{\bf Top panel}: X-ray band ratio curve of \transient, 
calculated as the ratio of the count rate in the 1.5$-$10 keV band (H)
to that in the 0.3$-$1.5 keV band (S).
{\bf Bottom panel:} hardness ratio curve (H-S)/(H+S).  
In both panels blue points are data taken in WT mode; 
red points are data in PC mode.
The vertical dashed green line marks the abrupt drop in the rate 
light curve.
\label{fig:HR}}
\end{figure}


The band ratio and hardness ratio curves of \transient\ are shown 
in Figure~\ref{fig:HR}.
We have defined the energy range 0.3$-$1.5 keV as the soft band (S) 
and the energy range 1.5$-$10 keV as the hard band (H).
We have calculated both the ratio of the counts detected in the
two bands (H/S) and the hardness ratio (H-S)/(H+S).
The curves have been binned in order to have at least 500 net counts 
in each band up to day 507 post-trigger.
For the last point (covering the time interval from day 508 to day 1096) 
net counts in the required bands were derived  
from the summed post-drop observations with the same procedure used for
the count rate light curve.  Errors have been estimated in the Poisson
approximation \citep{Gehrels86}
and propagated with standard error propagation formulae. 
At the beginning, 
both curves show variations tracking the light curve flares,
with the hardness rising when the average rate increases, 
and decaying when the average rate decreases.
But later phases with different trends can be identified.  
We observe a hardness plateau phase from $\sim$2.4 to $\sim$16 days 
post-trigger,
corresponding to a phase of average rate decrease (from $\sim$2.4 to $\sim$6 days)
followed by a rapid rise by a factor $\sim$10 in $\sim$0.3 days 
and a shallow decay afterwards.
After 16 days, while the average rate in the light curve goes on
decaying  steadily, 
we observe a relatively rapid ($\sim$1 day long) drop 
(i.e. softening) in the hardness followed by a steep rising 
(i.e. hardening) phase that lasts up to $\sim$100 days,
when a final hardness plateau phase starts.
In both cases the last point represents a clearly different 
and much softer spectral state compared to the final trend
of the curve.  
A comparable softness is attained only at a few light curve minima 
between flares (e.g. the first one at $\sim$0.2 days), 
which are minima for the band ratio and hardness ratio curves as well.
If the abrupt \transient\ emission drop on day 508 
is due to the turn-off of the jet \citep{Zauderer13_SwiftJ1644}, 
which might be likely caused by a transition to a thin disk 
as the accretion rate dropped below a critical value 
of several tens of percent of the Eddington accretion rate 
\citep{Zauderer13_SwiftJ1644,Tchekhovskoy14_SwiftJ1644},
then the low-level residual emission we detected later should
have a different origin and this discontinuity in the hardness
curves would not be surprising.


%
\begin{figure}[ht]
\resizebox{\hsize}{!}{\includegraphics[angle=0]{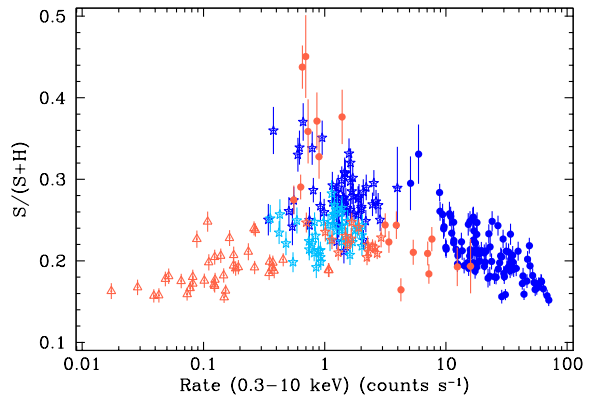} }
\caption[]{Soft band ratio S/(H+S) versus 0.3$-$10 keV rate for the light 
curve of \transient. 
Dark blue and light blue points are data taken in WT mode, 
the former already presented in \citetalias{Burrows11_SwiftJ1644} 
(up to 23 days since the trigger, Supplementary Figure~4), 
the latter (from 24 to 39.5 days) shown here for the first time. 
Red points are data in PC mode.
Filled circles represent WT and PC data up to 2.4 days since the trigger; 
empty stars represent WT and PC data between 2.4 and 39.5 days since the trigger; 
empty triangles represent PC data between 39.5 and 507 days post-trigger. 
\label{fig:HRvsRATE}}
\end{figure}


We see evidence of different spectral phases during \transient\ 
evolution also in Figure~\ref{fig:HRvsRATE}, where the soft band
ratio S/(H+S) is plotted against the total rate (H+S) in the 
0.3$-$10 keV band. Here XRT data in different operation mode 
are color-coded as in the previous figures, and different markers
distinguish early emission (before 2.4 days, filled circles)
from intermediate (between 2.4 and 39 days, empty stars) 
and late emission (after 39 days, empty triangles). 
According to this plot early data show an anti-correlation
between the soft band ratio and the overall count rate 
i.e. a trend for harder spectra at higher rates. 
The correlation is tighter for WT data. 
PC data are more dispersed and seem to lie on a different correlation, 
also consistent with the harder-when-brighter trend.
Late data are in PC mode only, and follow an opposite trend of harder
spectra at lower rates despite the large dispersion.
Intermediate data populate the central part of the plot.
Part of these (WT data up to 23 days, dark blue empty stars) 
were included in the corresponding plot in \citetalias{Burrows11_SwiftJ1644} 
(Supplementary Figure 4) and are interpreted as a second branch
of the WT harder-when-brighter correlation after a count-rate discontinuity
at $\sim$4~s$^{-1}$. The inclusion of the remaining WT data
(light blue empty stars)
has not completed the correlation in the left upward direction 
as expected, 
but simply thickened the data cloud on the left direction.
Note that count rates as low as those reached by the WT data
in Figure~\ref{fig:XRT_lc} are not seen in this plot because 
of the different binning criterion used. 
The intermediate PC data populate the bottom of the central cloud,
spanning a range in rates comparable to the intermediate WT data,
but with much less spread in softness (consistent with the plateau
observed in Figure~\ref{fig:HR} between 2.4 and 16 days).
With the possible exception of intermediate PC data, 
intermediate WT and late PC data in Figure~\ref{fig:HRvsRATE}
may be interpreted 
as a single component obtained by shifting leftwards 
and downwards with time an anti-correlation law similar to
the one followed by the early WT data.
This result could be obtained with a global hardening with time
superimposed on a strict hardness-tracking-brightness rule
on the smaller variability timescales of flares and dips.

\subsection{Spectral Analysis and Conversion to Flux}
\label{subsec:spectra}

The XRT team's on-line light curve repository 
\citep[\url{http://www.swift.ac.uk/xrt_curves/00450158/};][]{Evans09} 
uses a single conversion factor between count rate and flux (ECF), 
determined using an automated spectral fit of the PC mode data. 
This works well for GRBs, 
since there is rarely any significant spectral evolution after the first 
few ks of the afterglow, but it does not work well for \transient, 
due to the extremely strong spectral variations during 
the first few hundred days (Figure~\ref{fig:HR}).
A more realistic conversion to flux can be obtained via
time-dependent spectral analysis. 
The procedure, which we already applied in \citetalias{Burrows11_SwiftJ1644}, 
consists of the following steps: 
{\it i)} split the data set into a sequence of short time intervals 
tracking the hardness states of the source but with enough 
counting statistics to use standard $\chi^2$ fitting techniques, 
and extract spectra for each time interval; 
{\it ii)} fit each spectrum, derive the ECF corresponding to 
the best-fit model, and create a stepped ECF evolution law;
{\it iii)} convert the count rate light curve to flux 
by applying at each point an ECF obtained through 
a cubic spline interpolation of the ECF stepped curve.

The conversion applied in \citetalias{Burrows11_SwiftJ1644} 
was limited to the first 50 days of the \transient\ outburst, 
and was based on spectral modeling with the log-parabola model 
defined as
$ A(E) = E^{(-\alpha + \beta \log_{10}\left( E \right) )} $
where $\alpha$ is the photon index and $\beta$ is a measure
of the curvature compared to a simple power law, which is 
obtained for $\beta = 0$.
The complete spectral model we use also includes 
two absorption components: a Galactic one with
$\rm{N}_{\rm{Hgal}}$ fixed to 1.7 $\times$ 10$^{20}$ cm$^{-2}$ \citep{Kalberla05}
and an intrinsic one at redshift $z$ = 0.35 with \NH\
allowed to vary with time. 
The log-parabola model has been introduced as a good empirical 
fit for broadband spectral distributions of BL Lacs and other 
blazar sources. 
The version we use corresponds to the model {\tt logpar} in {\tt xspec} 
\citep{Massaro04_log-parabola}
with (constant) parameter pivot-$E$ (representing the 
low end of the energy range used in the fit) fixed to 1 keV, 
and curvature parameter equal to $-\beta$.  
In \citetalias{Burrows11_SwiftJ1644} we reported that the 
log-parabola spectral model generally
provided better fits than a simple power law 
to both time and intensity-selected spectra of \transient,
though power-law fits were statistically acceptable.
However, in about 74\% of cases 
the best-fit value of the parameter $\beta$  
was consistent with zero within its uncertainties, 
making the model perfectly
equivalent to a power law (see Supplementary Figure~11 in 
\citetalias{Burrows11_SwiftJ1644}), 
but with substantially larger errors on both 
the intrinsic absorption column and the photon index 
compared to the power-law fit.


%
\begin{figure}
\resizebox{\hsize}{!}{\includegraphics[angle=0]{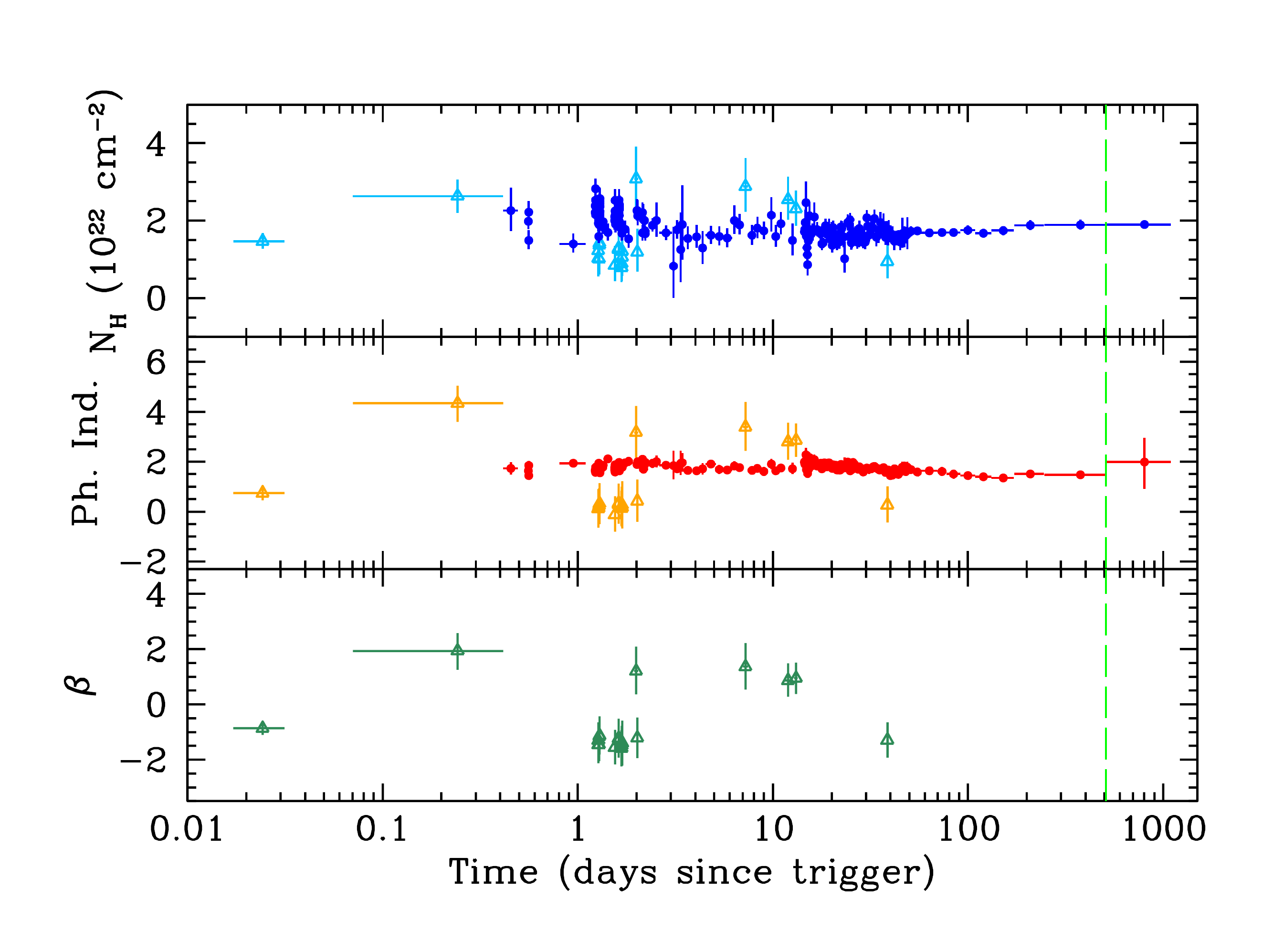} }
\caption[]{Temporal evolution of the best-fit parameters 
of the time-resolved spectra of \transient.
{\bf Top panel:} intrinsic absorption column.
{\bf Central panel:} photon index.
{\bf Bottom panel:} curvature of the log-parabola.
In all panels,
filled circles represent spectra well fit by an absorbed power-law model, 
while open triangles are used for the 20 spectra for which the absorbed 
log-parabola model gives a better fit according to F-test as described 
in the text. 
The vertical dashed green line marks the drop in the count rate light curve.
Remember that the photon index of the after-drop spectrum has been estimated
with intrinsic \NH\ value fixed at 1.9$\times$10$^{22}$ cm$^{-2}$.
\label{fig:param}}
\end{figure}


For the present paper,
we complete the set of time-dependent spectra 
used in \citetalias{Burrows11_SwiftJ1644}
with more spectra extracted from
the data collected after the first 50 days and
before the light curve drop, reaching a total of 
215 spectra (163 in WT mode and 52 in PC mode).
We have reanalyzed the whole data set with updated 
spectral response files, fitting each spectrum with
both power-law and log-parabola models\footnote{With 
the exception of the post-drop spectrum, where low 
statistics did not allow us to fit the log-parabola model.}. 
Since these two models are suitable for an F-test
\citep{Protassov02}, we apply it to estimate 
the probability P of chance improvement of the $\chi^2$
and reject all the log-parabola fits 
with P $>$ 0.01. This condition rejects 195 of
the log-parabola fits (including all the
best-fit solutions with $\beta$ consistent with zero
within errors) 
i.e., according to this selection 
the log-parabola model significantly improves the fit 
only for 20 spectra (15 in WT mode and 5 in PC mode).
Thus, for every further calculation, we decided to use 
the power-law fit results for all the spectra but the 20 
for which the log-parabola fit represents a significant 
improvement. 

We also extracted and analyzed a spectrum from the summed PC data
collected after the drop in the light curve, choosing source and 
background extraction regions as we did for the final count rate and
hardness ratio calculation.
In this case the data were binned to 1 count per energy bin to allow 
for fitting with Cash statistics \citep{Cash79} in {\tt xspec}
(with $\chi^2$ statistic test evaluation enabled). 
The very low statistics do not allow for a fit with 
an absorbed power-law with both 
intrinsic \NH\ and photon index free or 
an absorbed log-parabola model in this case.
We modeled the spectrum with an absorbed power-law with intrinsic 
\NH\ fixed to 1.9 $\times$ 10$^{22}$ cm$^{-2}$. 
The best fit photon index is $\Gamma = 2\pm1$. 
The fit is poor and the uncertainty is very large. 
$\Gamma$ values larger than 2 are likely to be preferred 
because corresponding models give H/S  $<$ 1.5. 
The correct band ratio may be achieved also with $\Gamma$ values 
smaller than 2 and lower intrinsic \NH.  
We then performed a second fit with $\Gamma$ fixed to 1 and free intrinsic \NH. 
The fit is apparently equivalent to the previous one and formally gives
\NH\ $< 0.5$ $\times$ 10$^{22}$ cm$^{-2}$, but the model cannot reproduce
an $H/S \sim$1 unless \NH\ is zero.



\begin{table*}[htp]
\begin{center}
\caption{\transient: Best Fit Parameters of Time-resolved Spectra. \label{tbl:param}}
\begin{tabular}{ccrrcccrr}
\tableline\tableline
 Number & XRT Mode & Start (days)\tablenotemark{a} & Stop (days)\tablenotemark{b} & \NH\ (cgs)\tablenotemark{c} & Photon Index\tablenotemark{d} & $\beta$ & $\chi^2$ & dof \\
\tableline
\\
1 & WT &   0.01713861 &   0.03135739 & 1.5$^{+0.2}_{-0.2}$ & 0.74$^{+0.30}_{-0.28}$ & -0.86$^{+0.25}_{-0.24}$ & 469.530 & 470 \\ 
2 & PC &   0.07028800 &   0.41441000 & 2.6$^{+0.4}_{-0.4}$ & 4.34$^{+0.71}_{-0.74}$ & 1.94$^{+0.65}_{-0.69}$ & 131.768 & 104 \\ 
3 & PC &   0.41559230 &   0.49312770 & 2.3$^{+0.6}_{-0.5}$ & 1.73$^{+0.25}_{-0.24}$ &  $-$  &  45.600 & 40 \\ 
4 & WT &   0.55755144 &   0.55925056 & 2.0$^{+0.2}_{-0.2}$ & 1.63$^{+0.10}_{-0.10}$ &  $-$  & 145.590 & 138 \\ 
216\tablenotemark{e} & PC & 508.39513700 & 1096.00273100 & 1.9$^{+0.0}_{-0.0}$ & 1.98$^{+0.97}_{-1.08}$ &  $-$  &  23.611 & {\bf 47} \\ 
\\
\tableline
\end{tabular}
\tablenotetext{1}{Starting time of data interval, relative to the first BAT trigger at $T_{0}$ $=$
2011 March 28 12:57:45.201 UTC $=$ 55648.5401 MJD.} 
\tablenotetext{2}{Ending time of data interval, relative to the first BAT trigger at $T_{0}$ $=$
2011 March 28 12:57:45.201 UTC $=$ 55648.5401 MJD.} 
\tablenotetext{3}{Units are 10$^{22}$ cm$^{-2}$.}
\tablenotetext{4}{Gives the value of the parameter $\alpha$ when the value of $\beta$ is listed.}
\tablenotetext{5}{Spectrum of the post-drop phase.}
\tablecomments{This table is published in its entirety in the 
electronic edition of the Astrophysical Journal.  
A portion is shown here for guidance regarding its form and content.}
\end{center}
\end{table*}


Our final best-fit results, including those for the post-drop spectrum, 
are listed in Table~1 
and shown in Figure~\ref{fig:param}.
The log-parabola model is a curved model
with a broad minimum in the spectrum 
at E$_{\rm crit} = 10^{(\alpha - 2) \frac 2\beta}$ keV if $\beta > 0$,
or a broad peak at energy E$_{\rm crit}$ for $\beta < 0$.
Note that among the log-parabola best-fit models 
that survive our selection, 
we have five with positive values of $\beta$
and all of them correspond to PC spectra.
In all cases E$_{\rm crit}$ is in the 2.5$-$6.0 keV range,
well within the \swift/XRT energy band, 
but the peak/minimum is so broad compared to the band
(relative width $\sim$~$ \left|\alpha / \beta\right| $) 
that almost no curvature can be seen by eye in the spectrum. 
Finally, the 20 log-parabola fits show a strong correlation 
between N$_{\rm H}$, photon index, and $\beta$. 
This is apparently unrelated to any physical status, because 
the 20 spectra do not show any common property 
(e.g. corresponding to peaks or minima, showing pile-up, etc.).
We are led to believe that in these 20 special cases 
the log-parabola fits are merely approximate descriptions
that we cannot use for any other purpose than flux estimation.


%
\begin{figure*}
\resizebox{\hsize}{!}{
\includegraphics[angle=0]{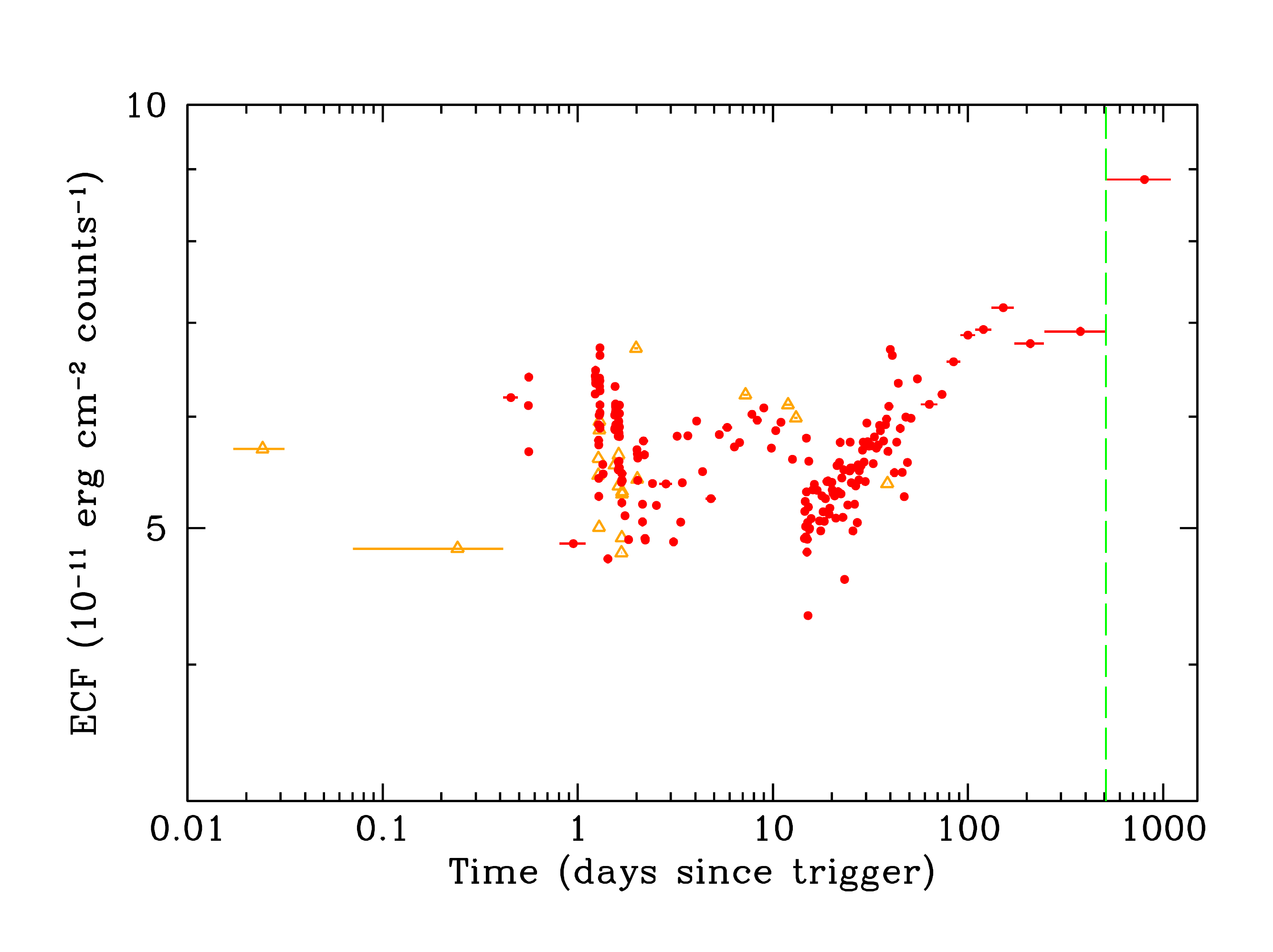}
\includegraphics[angle=0]{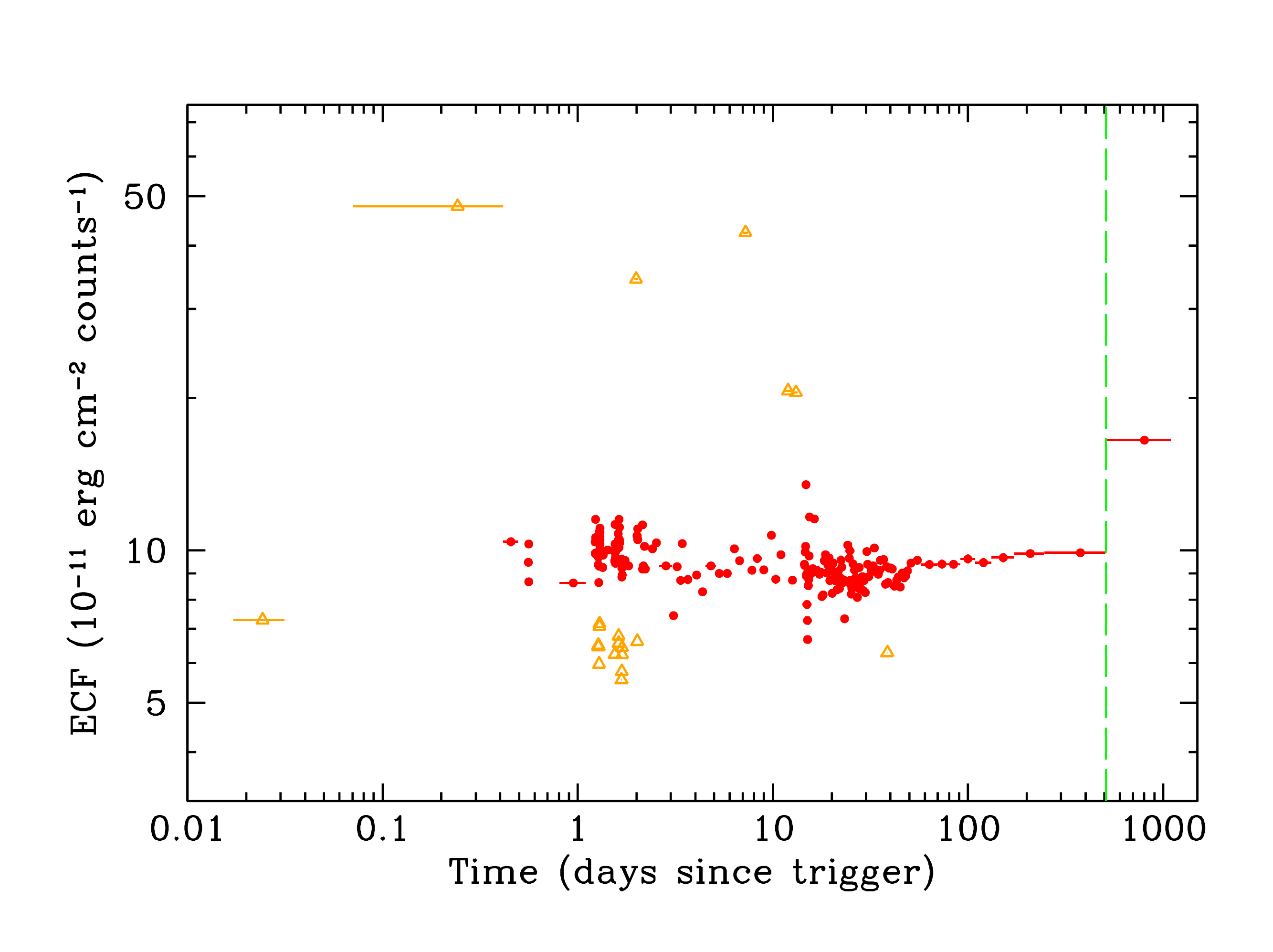} 
}
\caption[]{{\bf Left panel:} Energy Conversion Factors (ECFs) used 
to convert the count rate light curve in Figure~\ref{fig:XRT_lc} 
to observed flux units. 
Orange triangles represent ECFs coming from absorbed log-parabola fits,
and red circles represent ECFs coming from absorbed power-law fits.
The vertical dashed green line marks the drop in the rate light curve.
The variations in the ECFs mirror the strong variations 
in the band and hardness ratio plots of Figure~\ref{fig:HR}.
{\bf Right panel:} Energy Conversion Factors (ECFs) to be used
for conversion of the count-rate light curve of \transient\ into 
unabsorbed 0.3$-$10 keV flux. Symbols and colors are as in the left panel.
\label{fig:ECF}}
\end{figure*}


The ECFs to be used for the final conversion from count rate 
(Figure~\ref{fig:XRT_lc})
to 0.3$-$10 keV observed flux 
are derived from the best-fit results listed in Table~1 
and are shown in the left panel of Figure~\ref{fig:ECF}, 
while in the right panel of the same figure we show 
the corresponding ECFs for conversion into 0.3$-$10 keV unabsorbed flux.
Note that in this latter case 
the five log-parabola best-fit models with $\beta > 0$ 
lead to higher than average values of the ECF, and all those with
$\beta < 0$ 
lead to lower than average values of the ECFs.
This is due to log-parabola spectra with $\beta>0$
having larger 
than average \NH\ values and softer than average photon indices, 
as opposed to the $\beta<0$ 
solutions, for which the opposite is true.
The correction for the absorption leads to a larger scatter in the ECFs.
The absolute average deviation of the ECFs 
(i.e. the average of their differences from the mean value, taken in absolute value)
rises from 7\% to 15\% of the mean value going from the left panel to the right panel
in Figure~\ref{fig:ECF}.

The final light curve in observed and unabsorbed flux will be 
introduced and discussed in Section \ref{subsec:flux_lc}.
Note that the ECFs for the observed flux after day 15 change systematically, 
mirroring the strong variations in the band and hardness ratio plots of Figure~\ref{fig:HR}.
This could lead to a different slope for the overall decay rate than 
would be obtained using a single mean ECF.  
The ECFs for the unabsorbed flux, instead, show an average constant trend after day 15, 
in better agreement with the standard XRT light curve repository. 
%


%
\begin{figure*}
\resizebox{\hsize}{!}{
\includegraphics[angle=0]{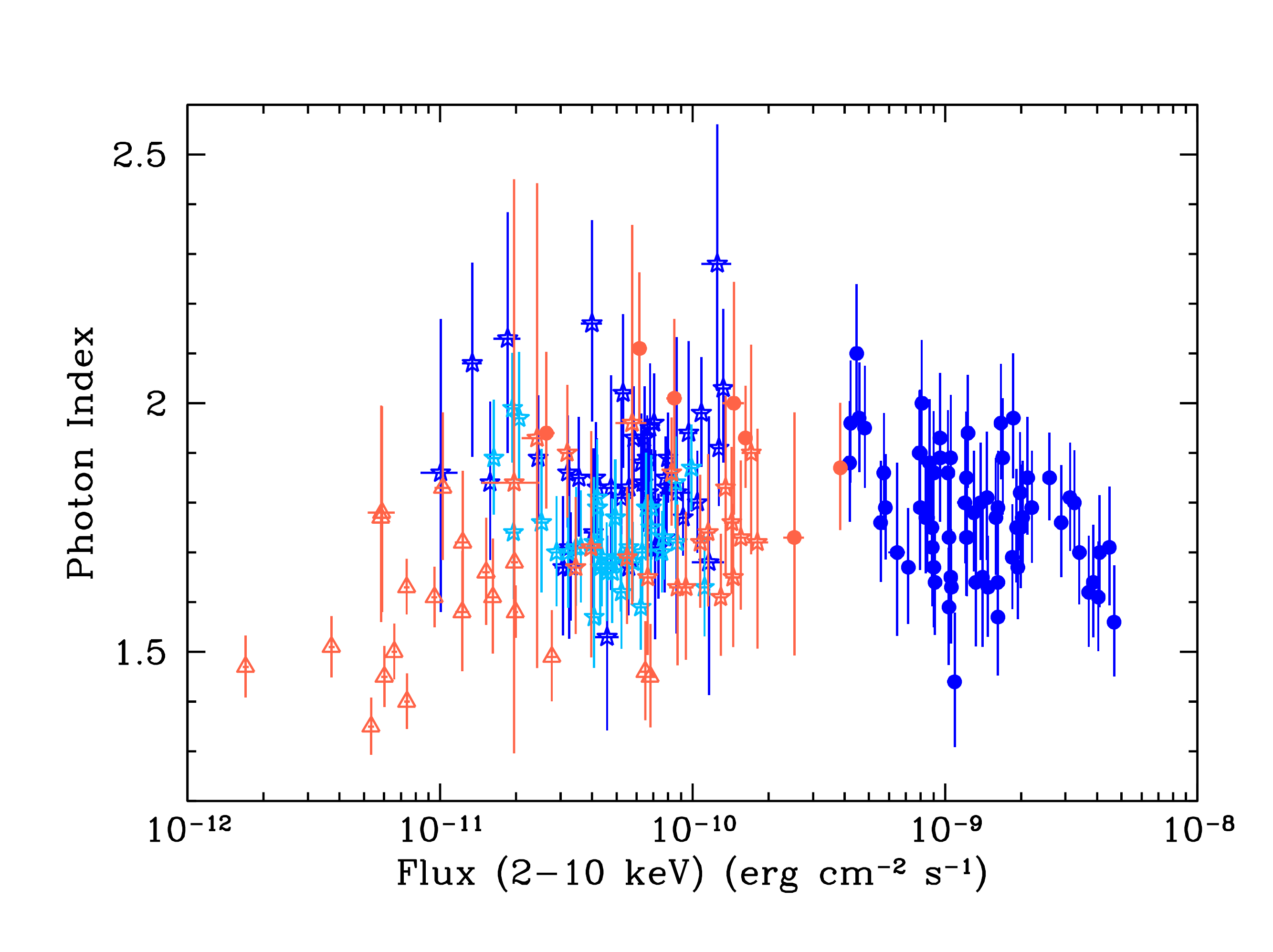} 
\includegraphics[angle=0]{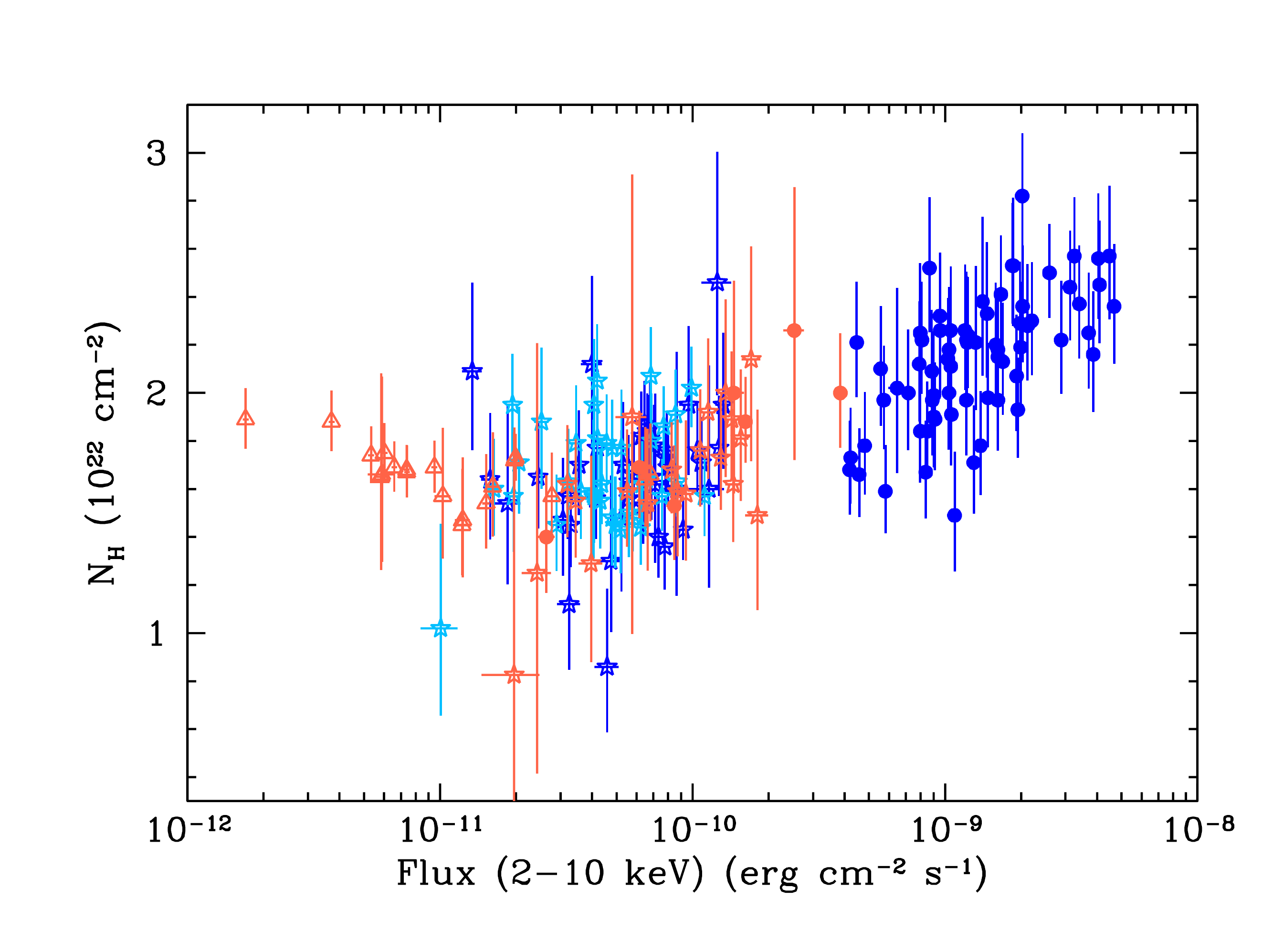} 
}
\caption[]{{\bf Left panel:} photon indices of the time-resolved spectra of \transient\
well fit by an absorbed power-law model as a function of the 2$-$10 keV flux. 
{\bf Right panel:} intrinsic absorption column \NH\ as function of the 2$-$10 keV flux
for the same spectra used in the left panel.
In both panels colors and symbols are defined as in Figure~\ref{fig:HRvsRATE}.
\label{fig:ParvsFlux}}
\end{figure*}


A closer look at the temporal evolution of the spectral parameters
in Figure~\ref{fig:param} shows that the variations of both 
the intrinsic \NH\ and the photon index approximately track the rate, 
but with opposite trends: the \NH\ tends to be higher at higher rates, 
while the photon index tends to be lower at higher rates, 
at least for the early portions of the light curve\footnote{Note 
that statistical correlation between \NH\ and photon index in fits 
goes in the direction of higher \NH\ at higher photon index,
and cannot be responsible for the observed trends with rate.}.
These different behaviors are visible in Figure~\ref{fig:ParvsFlux},
where the photon indices and the intrinsic absorption columns of the 
time-resolved spectra of \transient\ well fit by an absorbed power-law 
model are plotted as a function of the 2$-$10 keV flux. 
The photon index plot (left panel) shows a structure 
similar to the soft band ratio versus rate plot in Figure~\ref{fig:HRvsRATE},
though a larger dispersion in the different groups of data is clearly seen.
The intrinsic \NH\ plot (right panel) has a different structure,
with early data (filled circles) showing a more-absorbed-when-brighter trend,
and late data (empty triangles) suggesting a moderately increasing absorption
as the flux decays. The intermediate data (empty stars) are more difficult 
to interpret because of the large dispersion, 
but seem to follow a trend similar to early data.
A general trend of intrinsic \NH\ decreasing with time is confirmed 
by a linear fit of \NH\ versus $\log_{10} t$. With $t$ in units of days 
we obtain an intercept at 1 day of (1.98$\pm$0.02)$\times 10^{22}$~cm$^{-2}$
and a linear coefficient of -0.44$\pm$0.02, 
with a $\chi^2_r =$1.522 (213 dof).
An analogous fit for the photon index gives
an intercept at 1 day of 1.82$\pm$0.01 
and a linear coefficient of -0.40$\pm$0.01, 
with a $\chi^2_r =$1.885 (213 dof), and confirms the general trend 
for hardening of the spectra we already noted in Section \ref{subsec:hr_lc}.

\subsection{X-ray Flux Light Curve}
\label{subsec:flux_lc}


%
\begin{figure}[t]
\resizebox{\hsize}{!}{\includegraphics[width=7truecm,angle=0]{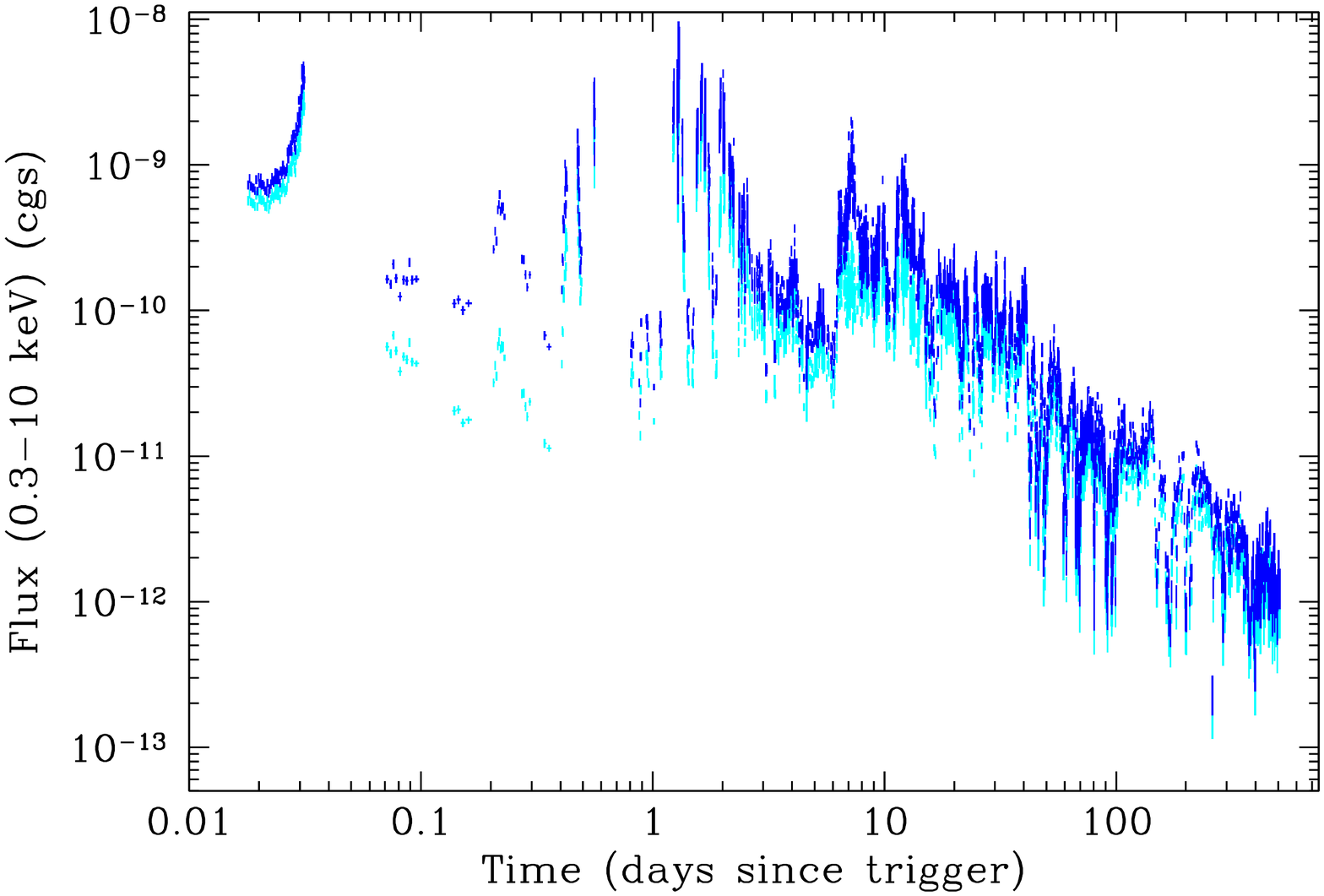} }
\resizebox{\hsize}{!}{\includegraphics[width=7truecm,angle=0]{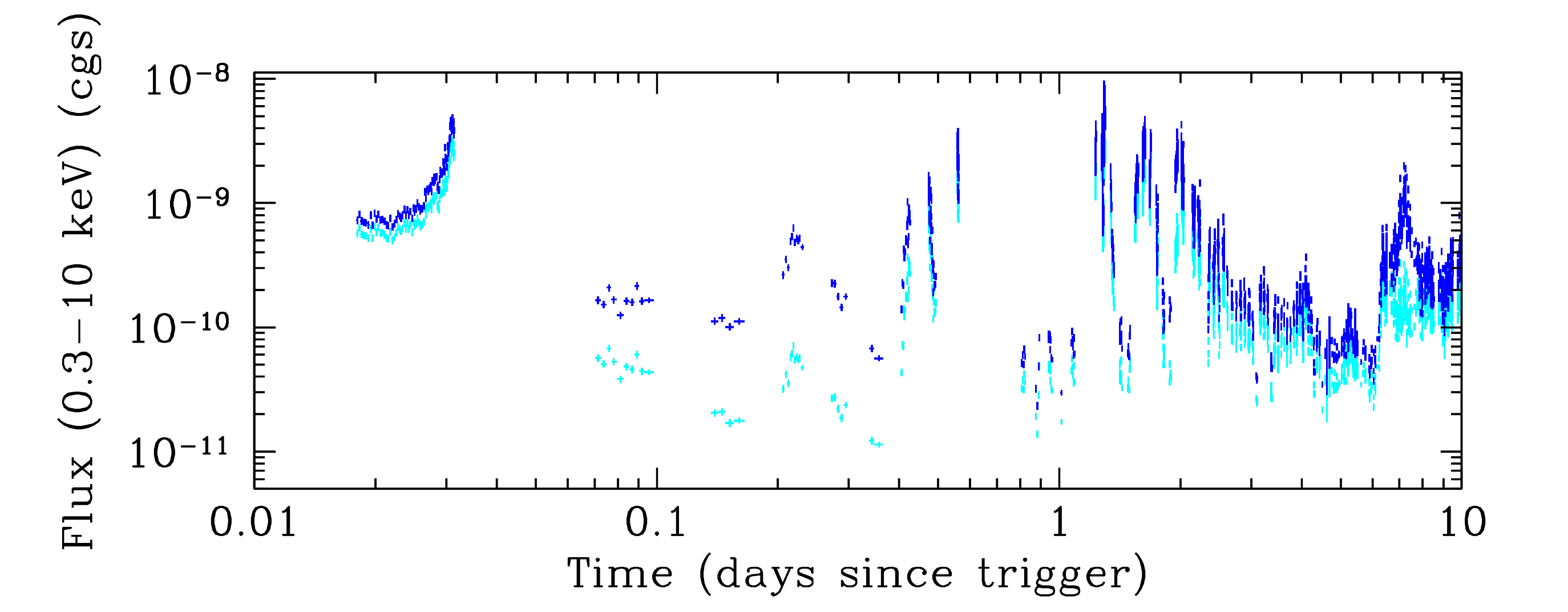} }
\caption[]{{\bf Top panel:} X-ray flux light curves of \transient\ in the 
0.3$-$10 keV band obtained with the time-dependent conversion procedure 
described in Section \ref{subsec:spectra} applied to the two sets of ECFs from 
Figure~\ref{fig:ECF}.  
The curve in observed flux is colored in light blue and the one in unabsorbed
flux is in blue.
{\bf Bottom panel:} detail of the first 10 days of the flux 
light curves of \transient. Color coding as in the top panel.
\label{fig:flux_lc_comp}}
\end{figure}



%
\begin{figure}[t]
\resizebox{\hsize}{!}{\includegraphics[width=7truecm,angle=0]{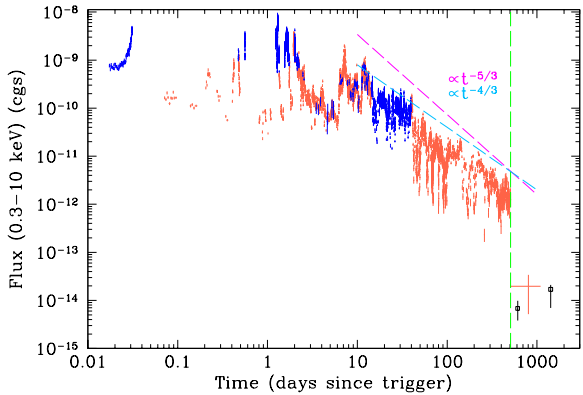} }
\caption[]{X-ray flux light curve of \transient\ in the 0.3$-$10 keV band, unabsorbed.
The XRT operation modes are color coded as in Figure~\ref{fig:XRT_lc}: 
blue for WT mode and red for PC mode.  
The vertical green dashed line marks the sudden drop of 2012 August 17.
The open black squares mark the two \chandra-ACIS observations performed
on 2012 November 26 and 2015 February 17 (see Section \ref{sec:obs} ).
For comparison, we show decay rates of $t^{-5/3}$ and $t^{-4/3}$ 
measured relative to the BAT trigger time.
\label{fig:flux_lc}}
\end{figure}



%
\begin{figure}[t]
\resizebox{\hsize}{!}{\includegraphics[angle=0]{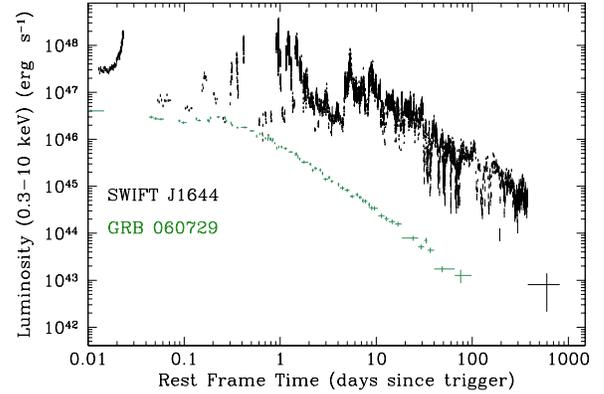} }
\caption[]{X-ray isotropic luminosity, \Lxiso, of \transient\ in the 0.3$-$10 keV band, 
compared to the isotropic luminosity of GRB\,060729, a long-lived but otherwise typical GRB afterglow at $z=0.54$.
No k-correction was performed, so the energy bands being compared are in the observed frame, 
but the difference in redshift is small (0.354 vs. 0.54).
\label{fig:lum_lc_vs_grb}}
\end{figure}



\begin{table}[htp]
\begin{center}
\caption{\transient\ X-Ray Light Curve. \label{tbl:flux_lc}}
\begin{tabular}{ccccc}
\tableline\tableline
t1\tablenotemark{a} & t2\tablenotemark{b} & Flux\tablenotemark{c} & Flux Uncertainty\tablenotemark{c} & XRT \\
(days)&(days)& (cgs)&(cgs)&Mode \\
\tableline
\\
  0.01720060 &   0.01742560 &  7.545e-10 &  5.449e-11  & WT \\ 
  0.01742561 &   0.01764019 &  7.904e-10 &  5.576e-11  & WT \\ 
  0.01764023 &   0.01790898 &  6.321e-10 &  4.634e-11  & WT \\ 
  0.01790898 &   0.01814023 &  7.300e-10 &  5.288e-11  & WT \\ 
  0.01814019 &   0.01835061 &  8.105e-10 &  5.702e-11  & WT \\ 
  0.07034943 &   0.07260637 &  1.648e-10 &  1.166e-11  & PC \\ 
  0.07260642 &   0.07512378 &  1.519e-10 &  1.074e-11  & PC \\ 
  0.07512371 &   0.07700449 &  2.083e-10 &  1.473e-11  & PC \\ 
  0.07700449 &   0.07940611 &  1.671e-10 &  1.181e-11  & PC \\ 
  0.07940616 &   0.08276264 &  1.249e-10 &  8.783e-12  & PC \\ 
\\
\tableline
\end{tabular}
\tablenotetext{1}{Starting time of data interval, relative to the first BAT trigger at $T_{0}$ $=$
2011 March 28 12:57:45.201 UTC $=$ 55648.5401 MJD.} 
\tablenotetext{2}{Ending time of data interval, relative to the first BAT trigger at $T_{0}$ $=$
2011 March 28 12:57:45.201 UTC $=$ 55648.5401 MJD.} 
\tablenotetext{3}{Units are erg cm$^{-2}$ s$^{-1}$.}
\tablecomments{This table is published in its entirety 
in the electronic edition of the Astrophysical Journal.  
A portion is shown here for guidance regarding its form and content.}
\end{center}
\end{table}


The final light curves of \transient\ in observed flux 
and unabsorbed flux are shown in Figure~\ref{fig:flux_lc_comp}.
The conversion procedure and the time-dependent ECFs used
are described in Section \ref{subsec:spectra}.
Note that discrepancies between the two curves are largest
in the time intervals
from 0.07 days after the BAT trigger to $\sim$0.5 days, and from 6 to 15 days.
The overall later decay does not seem dramatically affected 
by the correction for the absorption. 
Models of TDEs give predictions about the decay 
of the intrinsic luminosity of the source. 
For this reason the unabsorbed flux light curve of \transient\
should be used in testing models.
This one is shown alone in Figure~\ref{fig:flux_lc} 
and listed in Table~2. 

The X-ray light curve of \transient\ 
differs from the typical X-ray afterglow light curve 
of a GRB in many respects.
In Figure~\ref{fig:lum_lc_vs_grb} we compare the X-ray  
light curve of \transient\ to that of GRB\,060729, 
the longest GRB X-ray light curve ever obtained \citep{Grupe10}.

Both curves are shown in terms of the X-ray luminosity, \Lxiso, 
calculated assuming isotropic emission, 
and are plotted in the source rest frame.

The \Lxiso\ light curve of GRB\,060729 has 
a canonical afterglow shape, with a flat plateau phase followed by 
a power-law decay similar to the late average decay of \transient, 
but starting about 30 times earlier 
and running three orders of magnitude below the \transient\ light curve 
for $t > 10$ days. 
As illustrated by GRB\,060729, 
GRB afterglows are typically fairly smooth; they may show flares
(usually in the first day, before and/or during the plateau phase) but
never show dips, nor has a GRB ever exhibited a range of variability like that seen in \transient, which uniformly
spans more than one order of magnitude throughout 
its evolution.
Moreover, GRB afterglows rarely exhibit spectral evolution after the first day, while 
we have shown that \transient\ exhibits strong spectral evolution for well over a year.

The overall shape of the X-ray light curve of \transient\  
also differs from typical X-ray light curves from long-term 
monitoring of blazars, 
the class of AGN sources with observed radiation dominated 
by the emission from relativistic jets pointing at the observer.
The typical dynamic range of blazar flares in X-rays is a factor
of a few, not several orders of magnitude, and the typical activity 
duty cycle is small compared to that of the initial flares of \transient\
\citep{Krolik11_SwiftJ1644}.

\section{Light curve decay}
\label{sec:lcfit}

\begin{figure}
\resizebox{\hsize}{!}{\includegraphics[angle=0]{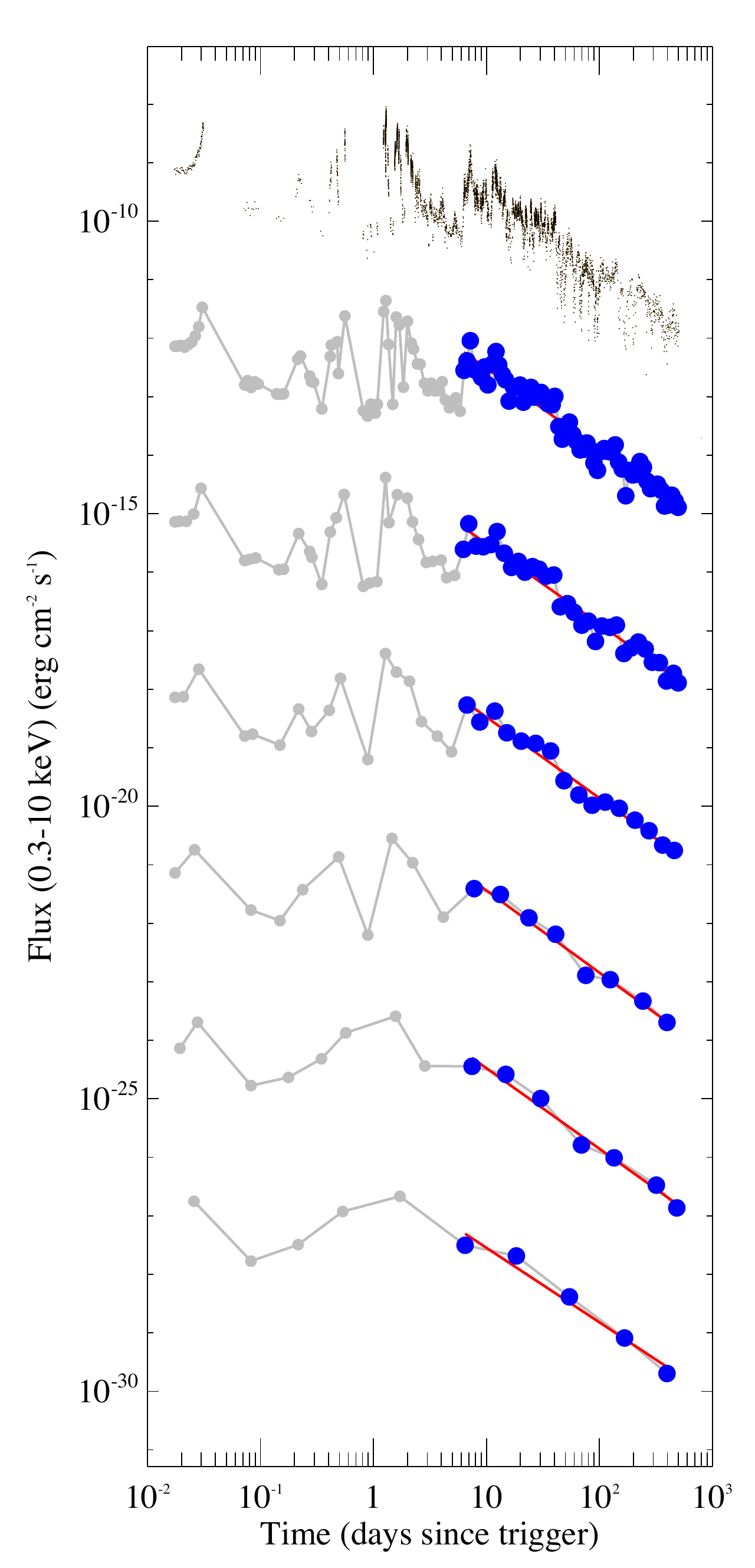} }
\caption[]{ 
Results of rebinning the unabsorbed flux light curve of \transient\ 
using $N$ time intervals per decade of days, 
equally spaced in $\log(t)$ as described in the text. 
From top to bottom, we plot the initial light curve 
and the six progressively smoothed light curves with 
32, 16, 8, 4, 3, and 2 points per decade, 
with arbitrary normalization. 
Points in the time interval 6$-$508 days, used for the fit, 
are highlighted with blue markers. 
Solid red lines represent the linear fits in the log$-$log space. 
\label{fig:smoothlc}}
\end{figure}

The X-ray light curve of \transient\ is highly variable
and affected by bright flares and deep dips at 
irregular intervals throughout the whole XRT monitoring,
but overall, there appears to be a continuous trend 
underlying the flaring and dipping activity in the light 
curve starting about 6 days after the trigger.
The upper envelope appears to follow a steady power-law decay 
from about $\sim$6 days post-trigger 
to the end of the outburst.
The modeling of this global trend is important 
for the interpretation of the phenomenon.
In order to do it, we proceeded as follows:

\noindent
{\it (1)} we generated a set of six progressively smoothed 
light curves by rebinning 
the unabsorbed flux light curve computed in Section \ref{subsec:flux_lc}
using a decreasing number of time intervals per decade of days, 
uniformly spaced in $\log(t)$, according to the sequence
32, 16, 8, 4, 3, and 2 intervals per decade (see Fig~\ref{fig:smoothlc}). 
For each time interval we calculated a rebinned point 
with time and flux coordinates obtained from 
the arithmetic averages of the logarithms of the times 
($\log_{10}\left( t \right)$) 
and the fluxes ($f_{\rm X}$) of all the data points therein. 
No error estimate through error propagation has been 
performed because errors of the single data points 
are negligible compared to the dispersion of the points.

\noindent
{\it (2)} we fit each smoothed light curve segment 
between 6 and 508 days post-trigger with a linear model 
in the $\log_{10}\left( t \right) - \log_{10}\left( f_{\rm X} \right)$ plane 
(i.e., $\log_{10}\left( f_{\rm X} \right)$ $=$ $ q - s\, \log_{10}\left( t \right)$)
and derived the best-fit value of the slope $s$ as
a function of the number of rebinned points.

\noindent
{\it (3)} we calculated the mean and sample variance 
of the six values of $s$ obtained in {\it (2)}
and used the result as a measure of the average slope of 
the flux light curve in the 6$-$508 days time interval.

We obtained a final decay slope of  $1.48 \pm 0.03$.
We checked the effect on the average decay slope 
of our time-dependent flux conversion 
by applying the same procedure (steps 1, 2, and 3 above)
to an unabsorbed flux light curve obtained 
with a uniform conversion factor
and to the flux light curve in the on-line repository as well.
We obtained values for the average decay slope in the
6$-$508 days time interval of $1.47 \pm 0.02$ and 
$1.45 \pm 0.05$, respectively. 
Both values are consistent within the errors with $1.48 \pm 0.03$.
This suggests that our time-dependent conversion to unabsorbed
flux does not affect the average decay slope of the light curve, 
though enhancing dispersion. Moreover, our light curve
allows for a more accurate estimate of the average decay slope 
compared to  the repository light curve. 

The average decay slope value we obtained
is definitely not consistent with the  
expected slope of \nicefrac{5}{3}.  The latter is considered a 
signature of tidal disruption because in the classical model by 
\citet{Rees88} and its updated versions \citep[e.g.,][]{Lodato09}
the decay law of the mass accretion rate  
resulting from fall-back of the disrupted stellar debris onto
the central BH is expected to be $\propto t^{-5/3}$, 
or at least asymptotically converging to this.
The same analysis applied to the observed flux light curve
has been presented in \citet{mangano_swift10proc}. In that case 
we obtained a final decay slope of $1.36 \pm 0.02$, which is 
consistent at the $2\sigma$ level with \nicefrac{4}{3}, i.e. the
slope value in the decay law for the jet luminosity in case
of super-Eddington slim accretion disk formation \citep{Cannizzo09}.
However, the average decay slope value of the unabsorbed flux light curve 
is also not consistent with \nicefrac{4}{3}. 
The measured slope may be consistent with the TDE phenomenon 
when more realistic physics of accretion disks 
is properly taken into account \citep{Cannizzo11_SwiftJ1644}.


%

\begin{figure}
\resizebox{\hsize}{!}{\includegraphics[angle=0]{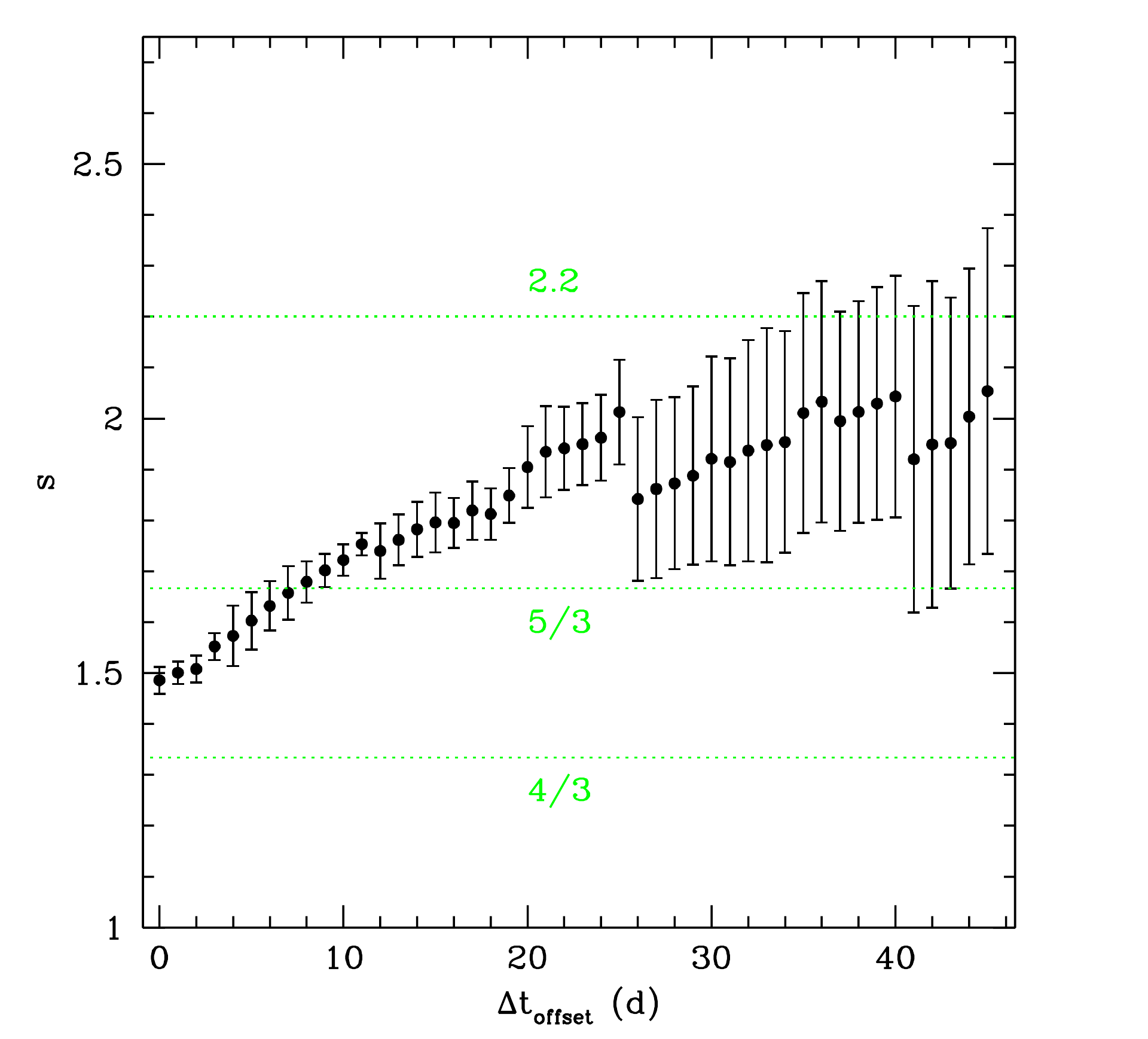} } 
\caption[]{Average  power-law decay slope of the flux light curve 
of \transient\ for $t> 6$ days post-trigger 
as a function of the offset of the tidal disruption event onset
from the BAT trigger. The horizontal lines at \nicefrac{4}{3}, \nicefrac{5}{3}, and $2.2$
represent expected values of the slope for different physical 
models of the disruption event. 
\label{fig:slope-offset}}
\end{figure}

Could it be that the actual TDE began much earlier than the nominal 
March 28 BAT trigger, and therefore the decay slope calculated 
from the true $t = 0$ was steeper?
We have investigated the possibility suggested by 
\citet{Tchekhovskoy14_SwiftJ1644} 
that the $t=0$ time appropriate for
the average light curve decay may lie
earlier than the BAT trigger time by 
$\Delta t_{\rm offset}$ days. 
For this purpose, we repeated the fits of our six smoothed 
light curves with the model 
$\log_{10}\left( f_{\rm X} \right) $ $=$ $ q - s\, \log_{10}\left( t+ \Delta t_{\rm offset} \right)$
and calculated the average best-fit slopes
for all integer values of $\Delta t_{\rm offset}$ from 1 to 45 days. 
The plot of the average slope $s$ versus $\Delta t_{\rm offset}$ 
is shown in Figure~\ref{fig:slope-offset}.
A slope consitent with \nicefrac{5}{3} at 1$\sigma$ 
is found for $\Delta t_{\rm offset} \sim$6~days.


%


   The degeneracy between decay steepness $s$ and  $\Delta t_{\rm offset}$
   revealed in Figure~\ref{fig:slope-offset}
   can be  quantitatively resolved
   by considering the form of the decay law,
   and the basic properties of the TDE decay.
      After a TDE occurs,
   there is negligible accretion from stellar fallback for $\sim$$t_{\rm fb}$,
   the fallback time for the most tightly bound debris to return
   to the BH and to begin to accrete.
   \citet{Tchekhovskoy14_SwiftJ1644}  suggest  that
   following  $t_{\rm fb}$ there may elapse
   an additional $\Delta t_{\rm offset}$ for magnetic flux accumulation
   in the inner disk to build up to the point that the jet
   becomes active.
   Therefore,
   one expects a gap of $t_{\rm fb}+\Delta t_{\rm offset}$, after
   which accretion begins,
     followed by a power law decay
    $\propto \left(t \over t_{\rm fb} +\Delta t_{\rm offset} \right)^{-s}$, 
     where $s\simeq 1-2$,  in the rate of supply of fallback gas
   to the central engine.
    However, the accretion-powered jet does not become visible
    until a time $t_{\rm fb} + \Delta t_{\rm offset}$ following the TDE.
    Thus in this idealization
 \begin{equation}
  \ f_{\rm X} = \left\{
    \begin{array}{l l}
             0      & \quad \text{  $t < t_{\rm fb}+ \Delta t_{\rm offset}$}\\
     f_{\rm X, max}\left( {t \over  t_{\rm fb} + \Delta t_{\rm offset} }\right)^{-s}
                 & \quad \text{   $t \ge t_{\rm fb}+ \Delta t_{\rm offset}$.}
    \end{array} \right.\
 \end{equation}
    Inherent in this formalism is the assumption that
 the jet luminosity $L_J$ tracks the accretion rate ${\dot M}$ from the inner
 accretion disk onto the BH for $t \ge t_{\rm fb}+ \Delta t_{\rm offset}$.
     The smooth power-law decay trend observed in the late XRT light curve
  supports the idea of a simple linear relation  
  between $L_J$ and ${\dot M}$
  out to $t\simeq 500$ days.

       By considering the ratio of the peak X-ray flux to the fluence
  $\Delta E_{\rm X, iso} = \int_{t_1}^{\infty} f_{\rm X}(t) dt$
   where $t_1 = t_{\rm fb}+ \Delta t_{\rm offset}$,
   one can directly measure $t_{\rm fb}+ \Delta t_{\rm offset}$ \citep{Gao12_SwiftJ1644} . 
    From the functional form for $f_{\rm X}(t)$,  one may write
 \begin{equation}
   t_{\rm fb} + \Delta t_{\rm offset} = (s-1) {\Delta E_{\rm X, iso} \over f_{\rm X, max}}
 \label{gaoeq}
 \end{equation}
    Note that since this argument depends only on the ratio of fluence
    and peak flux, all uncertainties such as beaming angle, accretion
    efficiency, jet efficiency, and distance
    cancel out.
  Using our measured values $ f_{\rm X, max}\simeq 9\times 10^{-9}$ erg cm$^{-2}$ s$^{-1}$
  and fluence $\Delta E_{\rm X} \simeq 6 \times 10^{-4}$ erg cm$^{-2}$ yields
       $t_{\rm fb}+ \Delta t_{\rm offset} \simeq 0.9$ days for $s=\nicefrac{5}{3}$,
   or  $t_{\rm fb}+ \Delta t_{\rm offset} \simeq 0.5 $ days for $s=\nicefrac{4}{3}$.
    These small values of $t_{\rm fb}+ \Delta t_{\rm offset} \la 1$ days argue against the
       possibility for $\Delta t_{\rm offset} \ga 10$ days
       presented in \citet{Tchekhovskoy14_SwiftJ1644}. 
    For this estimate we have idealized the entire light curve with
    one decay law. One could in theory fit a broken power law
    and obtain a more precise estimate, but given that
    $f_{\rm X, max}$ is set by a flare during the
   first few days, only a rough estimate seems warranted.

\section{Light curve dips}
\label{sec:dips}

Clues about the physical nature of \transient\ 
may come from the difference in the spectral distribution
of the radiation emitted by the source during the dips and
outside them. 
Since dips are short and often drop more than an order of magnitude 
from the average emission level, we cannot obtain sufficient counts 
for detailed spectral analysis of individual dips.
Instead, we must extract 
a single cumulative spectrum from many dips at a time. 
On the other hand, we cannot merge data from different operation modes.
Therefore, we consider the following two different periods 
during which significantly deep dips are visible in the
light curve of \transient:
{\it (a)} from $T_0$+14.5 days to $T_0$+39.5, with XRT steadily
observing in WT mode and 
{\it (b)} from $T_0$+40.5 days to $T_0$+404.8 with XRT steadily
observing in PC mode.

To select the dip good time intervals (GTIs),  
we developed a three step procedure 
that can be described as follows: 
{\it (i)} fit the light curve in a given time interval
to a model and calculate the weighted residuals;
{\it (ii)} remove all points with weighted residuals 
lower than a fixed negative threshold, then repeat the fit 
until no further points need to be removed;
{\it (iii)} shift the best-fit model down by a
proper amount and save as GTIs of
the dips all the time intervals including only points 
below the shifted model. 
We developed this procedure in the {\tt IDL} programming 
language and making use of the fit routines in the {\tt MPFIT} 
package by \citet{Markwardt09}. 

\begin{figure}
\vspace{-1.0truecm}
\hspace{-1.0truecm}
\resizebox{1.2\hsize}{!}{\includegraphics[angle=-90]{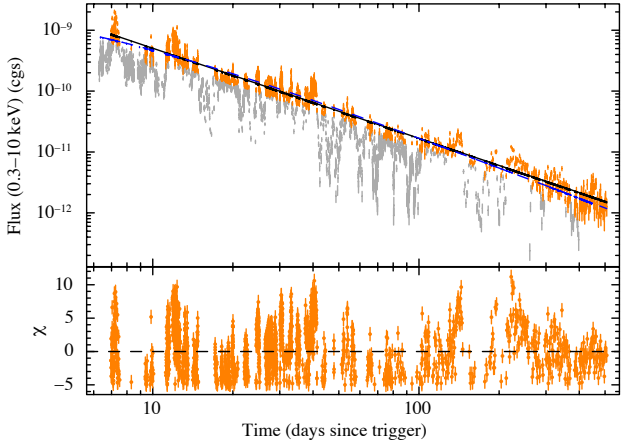} }
\vspace{-0.5truecm}
\caption[]{Fit to the X-ray flux light curve of \transient\  
over the time interval from $T_0$+6.3 days to $T_0$+507 days
performed with the iterative data rejection method 
described in the text.
The black solid line in the upper panel represents 
the best-fit solution obtained by fitting a power law model 
with free normalization and slope fixed to $1.48$ 
(expected value from estimates in Section \ref{sec:lcfit}).
All points with weighted residuals $< -5$ have been removed 
from the fit at each successive iteration. 
The final sets of rejected points (1833 out of 3740) 
are shown in the upper panel only, in light gray. 
The dashed blue line in the upper panel 
is the best-fit solution obtained with
a power-law model with initial time shifted by 
$\Delta t_{\rm offset}$ and slope fixed to \nicefrac{5}{3}. 
The value of $\Delta t_{\rm offset}$ derived after rejecting 
1894 points with weighted residuals $< -5$ for this model
is 5 days, in agreement with the plot in Figure~\ref{fig:slope-offset}.
\label{fig:lcfit}}
\end{figure}

Step {\it (ii)} in our procedure gives us an alternative way 
to estimate the continuum underlying the light curve, 
though because of the large dispersion of the data 
the reduced $\chi^2$ of the fit is not expected 
to assume statistically acceptable values, and the
uncertainties of the model parameters are expected
to be substantially underestimated. 
The dips represent the points with the largest deviation
from the continuum and they dominate the residuals: 
best-fit solutions obtained without removing any point 
have normalizations that make them lie significantly below 
the bulk of the data set. 
The iterative removal of points with negative residuals
allows us to obtain solutions that actually pass through 
the most densely populated region of the curve.
A threshold of $-5$ for the weighted residuals has been chosen in order to obtain 
a good description of the flux light curve after 6 days, 
using a power-law model $ f_{\rm X}(t) = N t^{-s}$ 
with decay slope $s$ fixed to the value of $1.48$ 
estimated in Section \ref{sec:lcfit}. 
In Figure~\ref{fig:lcfit}, the model just described is represented 
by the black solid line in the upper panel, and the weighted residuals in the
lower panel refer to it. 
They suggest that the true underlying continuum is reasonably well reproduced.
An equivalent
description of the continuum is 
provided by the dashed blue line in the upper panel,
representing a best-fit solution obtained with
a power-law model with initial time shifted by 
$\Delta t_{\rm offset}$ days before the BAT trigger
($f_{\rm X}(t) = N \left(t+\Delta t_{\rm offset}\right)^{-s} $) 
and slope $s$ fixed. However, the very same relation between 
$s$ and $\Delta t_{\rm offset}$ already shown in Figure~\ref{fig:slope-offset} 
affects results obtained with this model.
From these analyses we infer that -5 is a reasonable value for 
the threshold of rejection of points with negative residuals.


%



\begin{table}[htp]
\begin{center}
\caption{\transient: GTI of the Dips \label{tbl:dips}}
\begin{tabular}{cccc}
\tableline\tableline
Number  & t1\tablenotemark{a} &  t2\tablenotemark{b} & XRT \\
&(days)&(days)&Mode\\
\tableline
\\
1  &  14.87580058  &  15.28213427  & WT  \\
2  &  15.28650302  &  15.29033219  & WT  \\
3  &  15.34551102  &  15.34946728  & WT  \\
35  &  41.84660666  &  42.70801067  & PC  \\
36  &  42.78064206  &  43.23784472  & PC  \\
37  &  43.70705003  &  43.82982205  & PC  \\
\\
\tableline
\end{tabular}
\tablenotetext{1}{Starting time of data interval, relative to the first BAT trigger at $T_{0}$ $=$
2011 March 28 12:57:45.201 UTC $=$ 55648.5401 MJD.} 
\tablenotetext{2}{Ending time of data interval, relative to the first BAT trigger at $T_{0}$ $=$
2011 March 28 12:57:45.201 UTC $=$ 55648.5401 MJD.} 
\tablecomments{This table is published in its entirety in the 
electronic edition of the Astrophysical Journal.  
A portion is shown here for guidance regarding its form and content.}
\end{center}
\end{table}

\begin{figure}
\resizebox{\hsize}{!}{\includegraphics[angle=0]{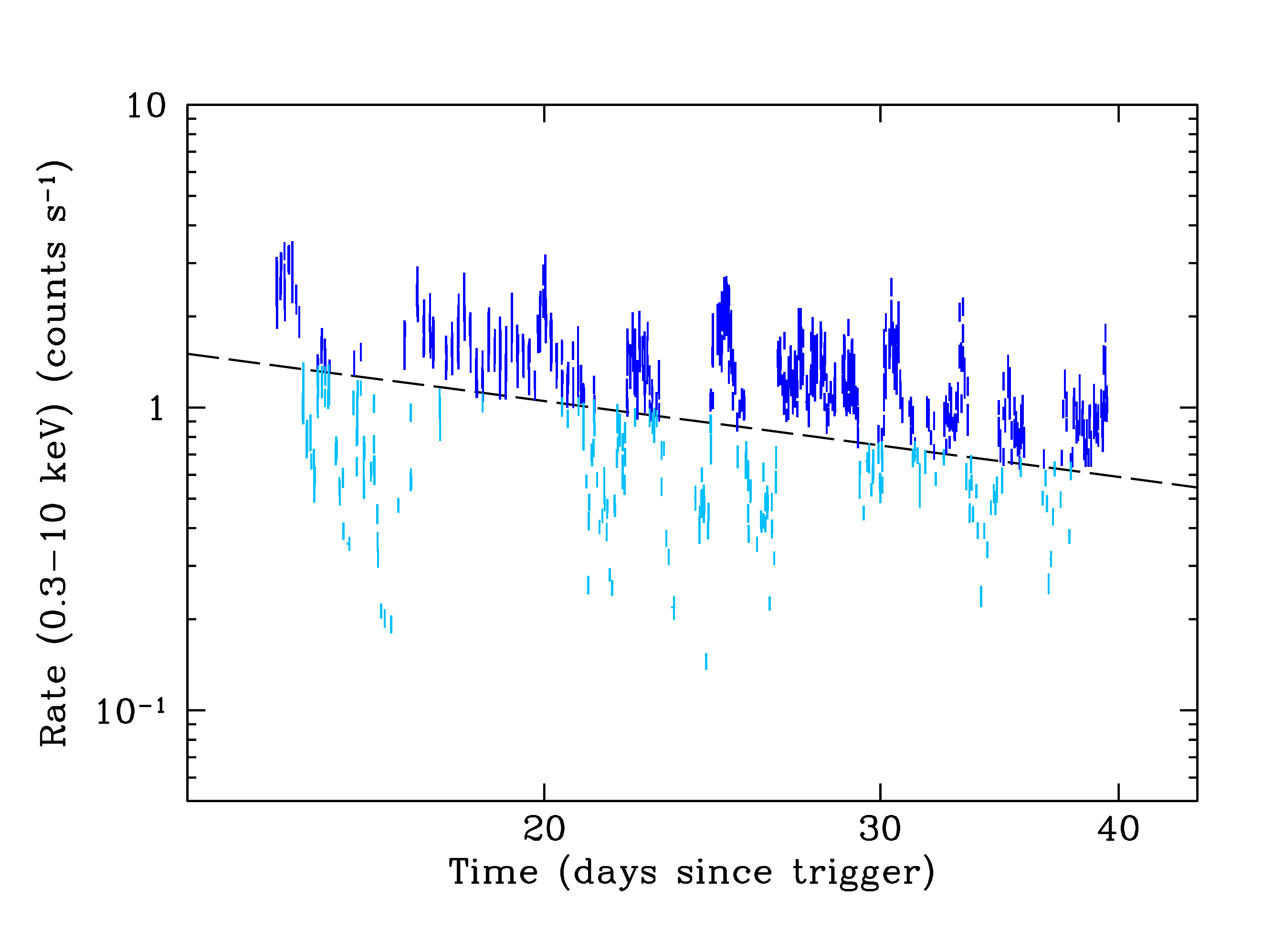} }
\resizebox{\hsize}{!}{\includegraphics[angle=0]{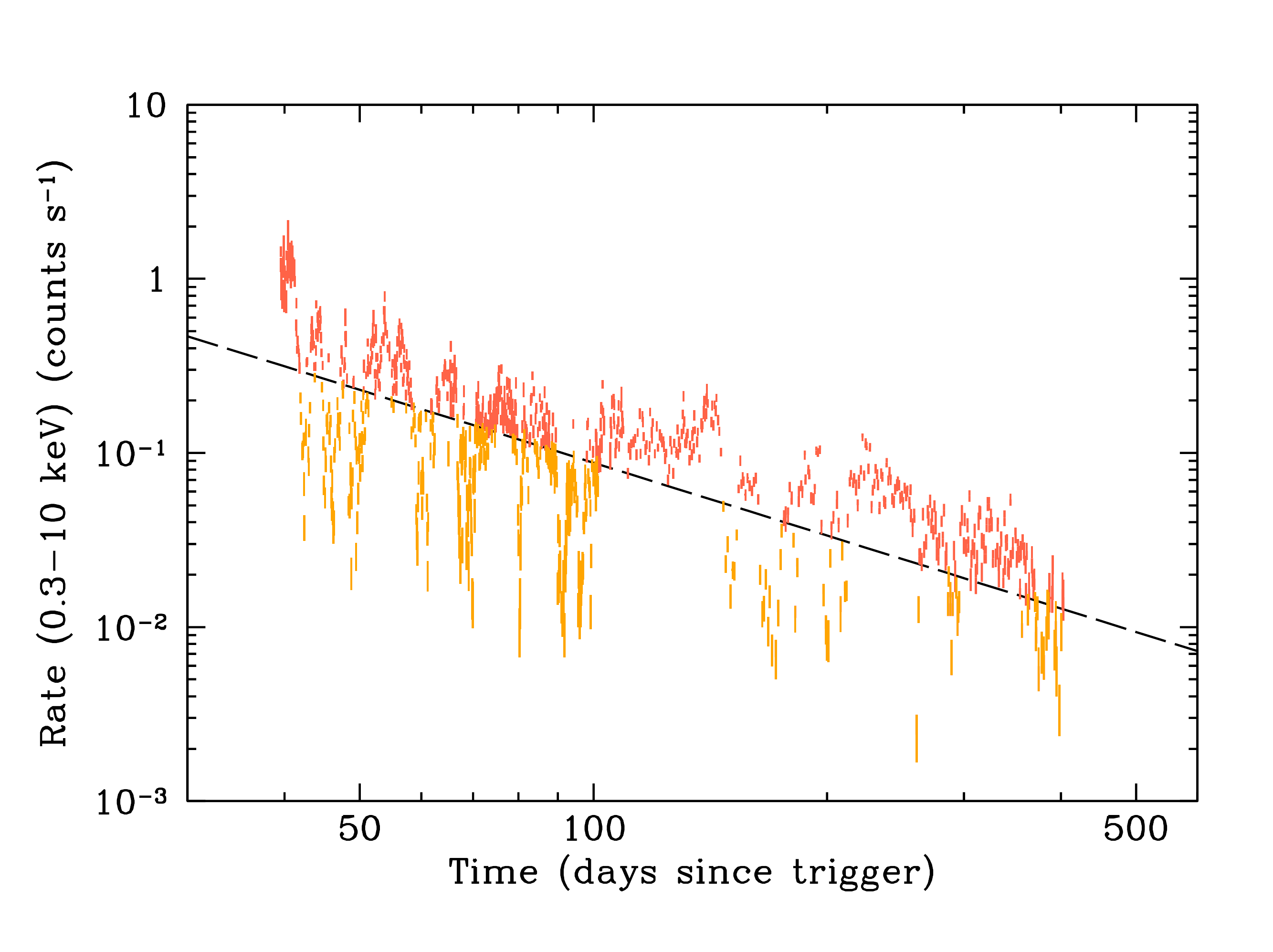} }
\caption[]{X-ray light curves showing the regions we identify as dips.
{\bf Top panel:} dips found during the observation period 
from $T_0$+14.5 days to $T_0$+39.5 days  (period {\it (a)} in the text) 
in WT mode. 
{\bf Bottom panel:} dips found from $T_0$+40.5 days to $T_0$+404.8 days 
(period {\it (b)} in the text) in PC mode. 
In both panels, the dashed line represents the cut between ``normal'' 
(above the line) and ``dip'' (below the line) time intervals, 
obtained as described in the text.
\label{fig:cut}}
\end{figure}
\begin{figure}
\resizebox{\hsize}{!}{\includegraphics[angle=0]{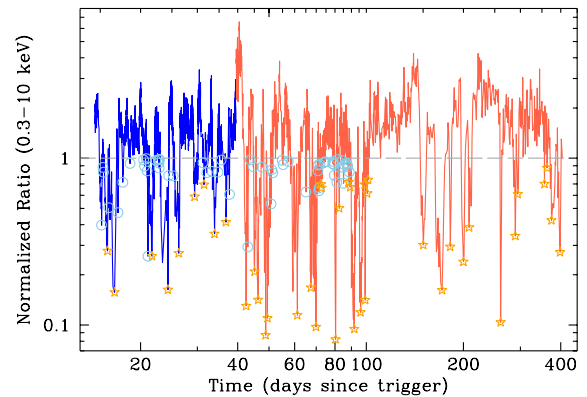} }
\caption[]{ Composite light curve obtained dividing the light curves 
in Figure~\ref{fig:cut} by their cutting lines. Blue data (before $T_0+$39.5 days)
are in WT mode and red data (after $T_0+$39.5 days) are in PC mode. 
The horizontal grey dashed line at level 1 is the new global cutting line 
defining the dips. Orange stars indicate the minima of the long ($>$0.5 days) dips
and light blue empty circles the minima of the short dips.
\label{fig:dips}}
\end{figure}

We applied our procedure to the count rate light curve 
of \transient\ computed in Section \ref{sec:redu},
separately for periods {\it (a)} (WT mode) and {\it (b)} (PC mode).
We used a power-law model to fit the underlying continuum,
and a threshold of -5 in step {\it (ii)}.
We obtained power-law decays with slopes $0.83 \pm 0.08$\footnote{ 
Errors on best-fit parameters given by {\tt MPFIT} 
are at 1$\sigma$ confidence, but in this case they are likely
underestimated by at least an order of magnitude.}
for period {\it (a)} and $1.390 \pm 0.005$ for period {\it (b)}, 
respectively.
Finally, we shifted each power-law model down by decreasing its
normalization by a quantity equal to 5 times the average percentage
error of the light curve points in the corresponding period, 
and defined the GTIs of the dips in each period as in step {\it (iii)}. 
We merged together successive time intervals containing only points
below the cutting line whenever they were not separated 
by at least one point at a rate 10\% larger than the cutting line.
The final time selection, consisting of 98 GTIs
(34 in WT mode and 64 in PC mode), is listed in Table~3. 
Figure~\ref{fig:cut} shows the separation between the ``normal'' and ``dip''
time intervals for the WT and PC mode data.
Figure~\ref{fig:dips} shows a composite normalized light curve
obtained by dividing each segment of the rate light curve 
by the corresponding cutting line. 
In this representation, the dips are the points below 1, and the
depth of each dip is measured by the normalized rate level of 
its minimum, the inverse of which tells us by what factor the
point is below the cutting line. In Figure~\ref{fig:dips} the
dips' minima are marked with an empty star if the corresponding
GTI is longer than 0.5 days, or with an empty circle if it is shorter.
Note that the durations of our GTIs span from $\sim$190~s
to $\sim$16.5 days, with 41 of them longer than 0.5 days and 57 shorter.
Moreover, the short GTIs are all concentrated in the first 90 days
and are not deeper than 0.2, 
with an average depth of $\sim$0.8,
while the long GTIs are uniformly distributed, with depths in the
range 0.87$-$0.08 and average depth $\sim$0.36.
%
%
It is likely that the absence of short GTIs after 90 days 
is a side-effect of the differential binning of the light curve, 
requiring longer time bins at later times, and average bin duration
larger than 0.1 days after the first 100 days.
If we define the dynamic range of a dip to be the ratio between 
the maximum count rate since the end of the preceding dip  
and the minimum count rate of the dip in question, we see that the short
dips have a small dynamic range (between 1.1 and 6.3 with
an average of 2.3), while the long dips have a much larger
dynamic range (between 2 and 54 with an average of 10).


%

%



\begin{table*}[htp]
\begin{center}
\caption{\transient: Spectral Analysis in and Outside Dips. \label{tbl:dips}}
\begin{tabular}{ccrrrcccrrc}
\tableline\tableline
 Type  & XRT Mode & Exposure (ks) & \NH\ (cgs)\tablenotemark{a} & Photon Index\tablenotemark{b} & $\beta$ & $\chi^2$ & dof & Flux (cgs)\tablenotemark{c}\\
\tableline
\\
normal  & WT  &  177.705  & 1.42$^{+0.05}_{-0.05}$   &  1.37$^{+0.08}_{-0.08}$  &  $-$0.36$^{+0.07}_{-0.07}$  &  942.714  &   791   &  71.75$^{+0.48}_{-0.56}$  \\ 
dips      & WT  &  117.681  & 1.43$^{+0.11}_{-0.11}$   &  1.50$^{+0.17}_{-0.17}$  &  $-$0.32$^{+0.15}_{-0.15}$  &  641.985  &   627   &  26.77$^{+0.34}_{-0.46}$  \\ 
normal  & PC  & 1074.540  & 1.84$^{+0.04}_{-0.04}$   &  1.51$^{+0.02}_{-0.02}$  &  $-$  &  802.450  & 743  &   7.18$^{+0.06}_{-0.06}$ \\ 
dips      & PC  &   325.291  & 1.78$^{+0.10}_{-0.10}$   &  1.69$^{+0.05}_{-0.05}$  &  $-$  &  359.828  & 383  &   2.80$^{+0.06}_{-0.07}$ \\ 
\\
\tableline
\end{tabular}
\tablenotetext{1}{Units are 10$^{22}$ cm$^{-2}$.}
\tablenotetext{2}{Gives the value of the parameter $\alpha$ of the log-parabola model when the value of $\beta$ is listed.}
\tablenotetext{3}{Observed flux in the 0.3$-$10 keV band in units of 10$^{-12}$ erg cm$^{-2}$ s$^{-1}$.}
\end{center}
\end{table*}

\begin{figure}
\resizebox{\hsize}{!}{\includegraphics[angle=-90]{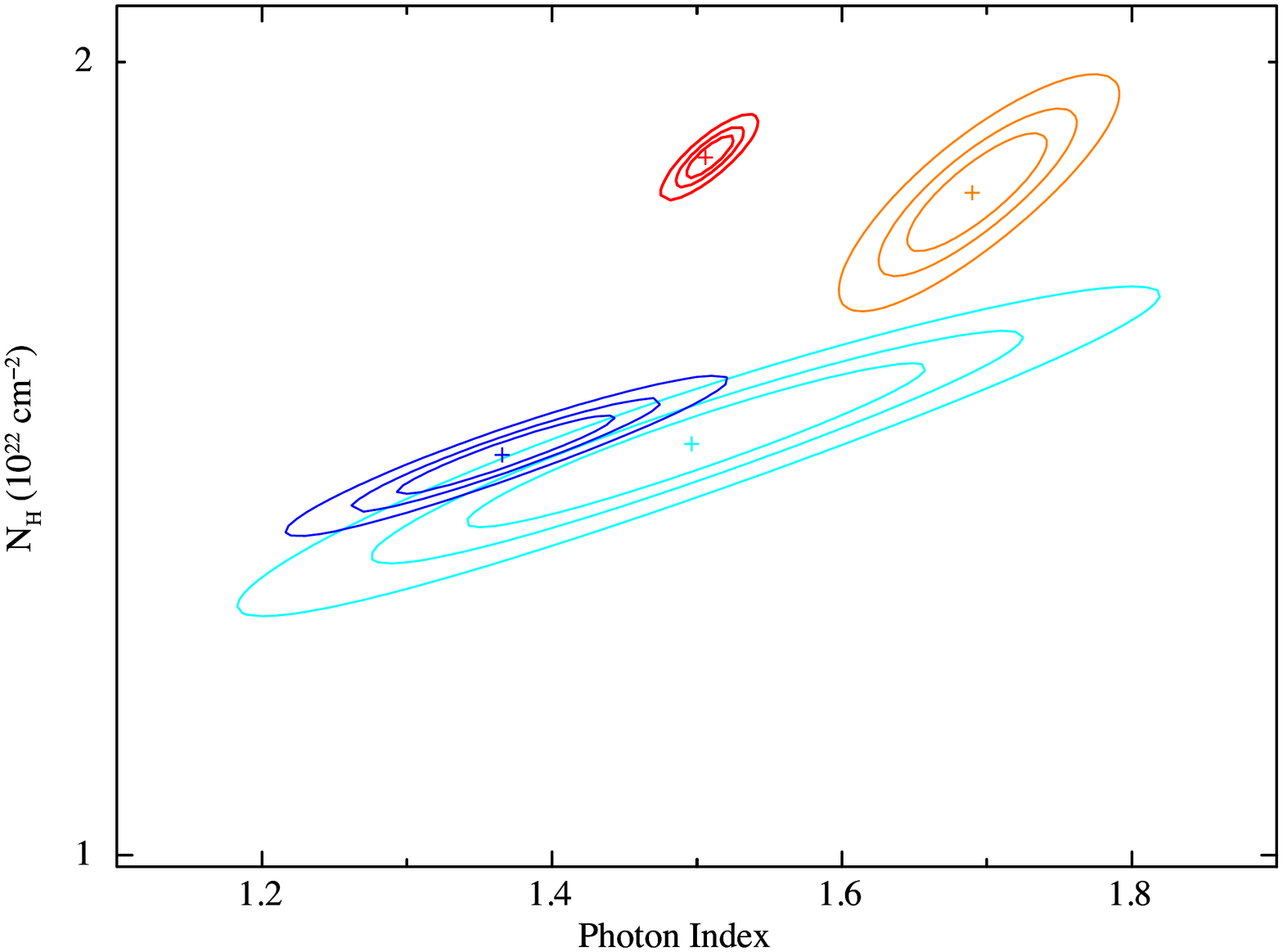} }
\resizebox{\hsize}{!}{\includegraphics[angle=-90]{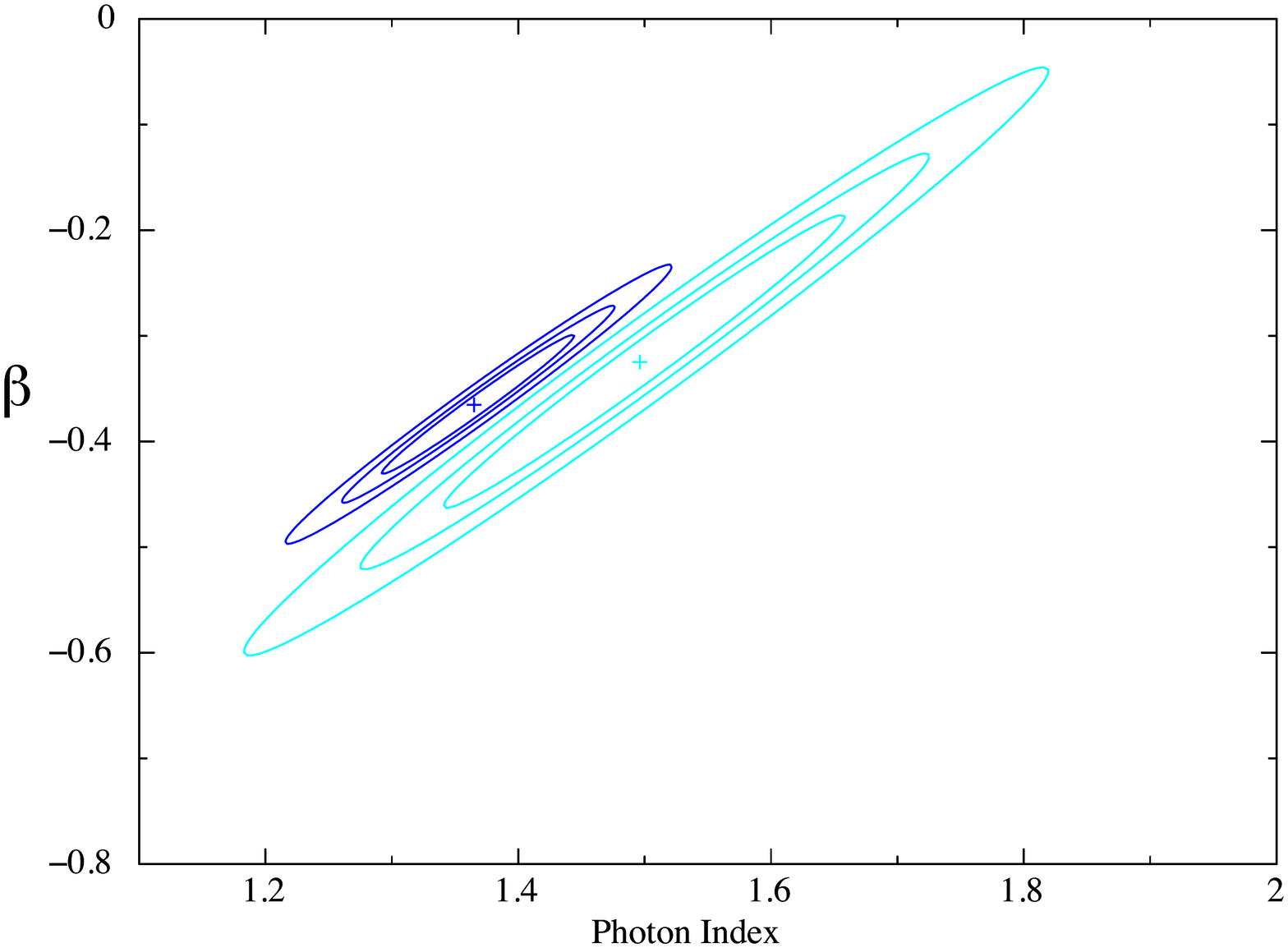} }
\caption[]{
{\bf Top panel:} contours at the 68\%, 90\%, and 99\% confidence level 
for \NH\ versus photon index, 
obtained from the fits of ``normal'' and ``dip'' spectra in WT and PC mode
(see results in Table~4, color coding is as in Figure~\ref{fig:spectra}). 
{\bf Bottom panel: } as in the top panel, for $\beta$ versus photon index.
\label{fig:contours}}
\end{figure}
\begin{figure}
\resizebox{\hsize}{!}{\includegraphics[angle=-90]{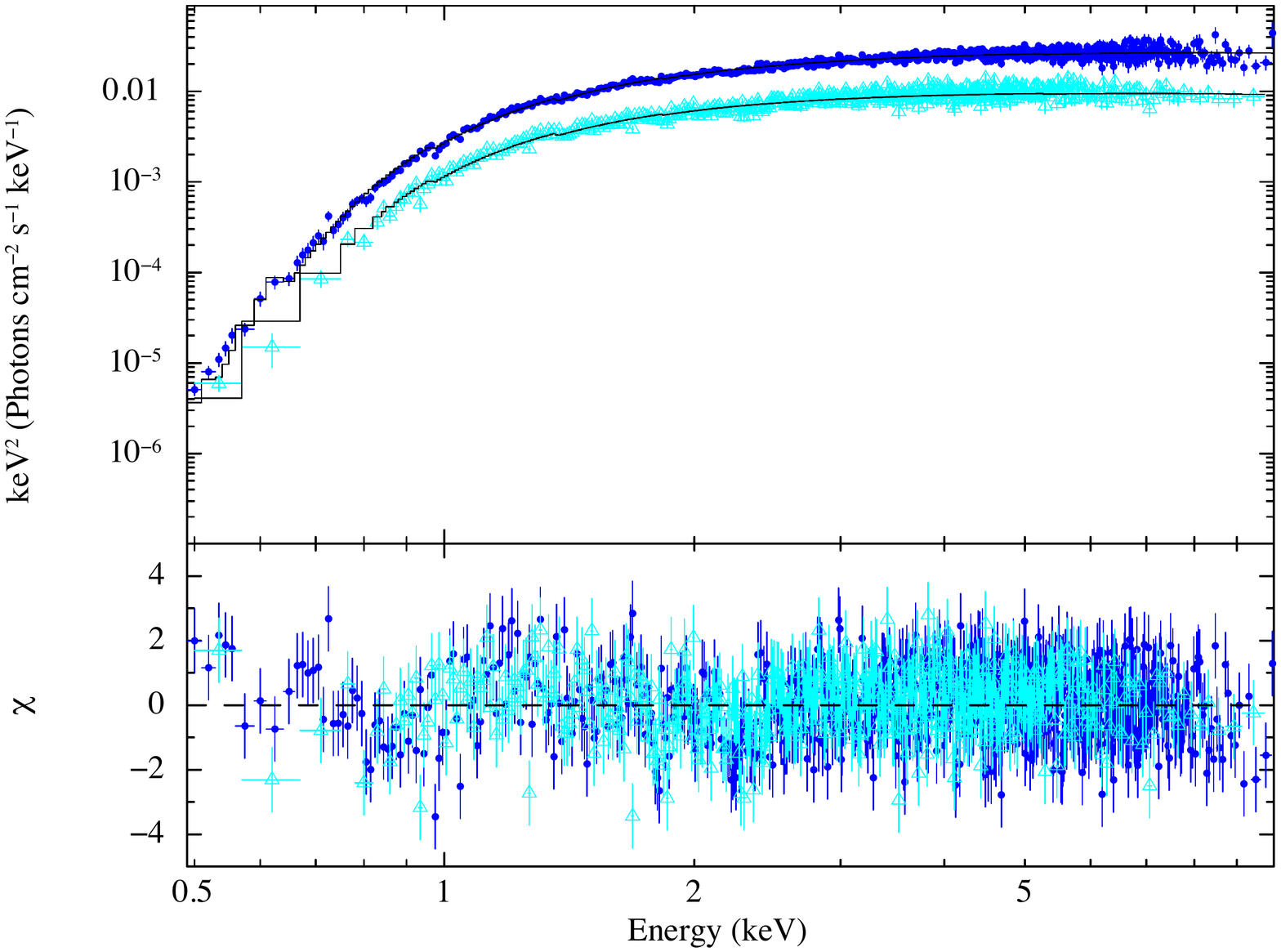} }
\resizebox{\hsize}{!}{\includegraphics[angle=-90]{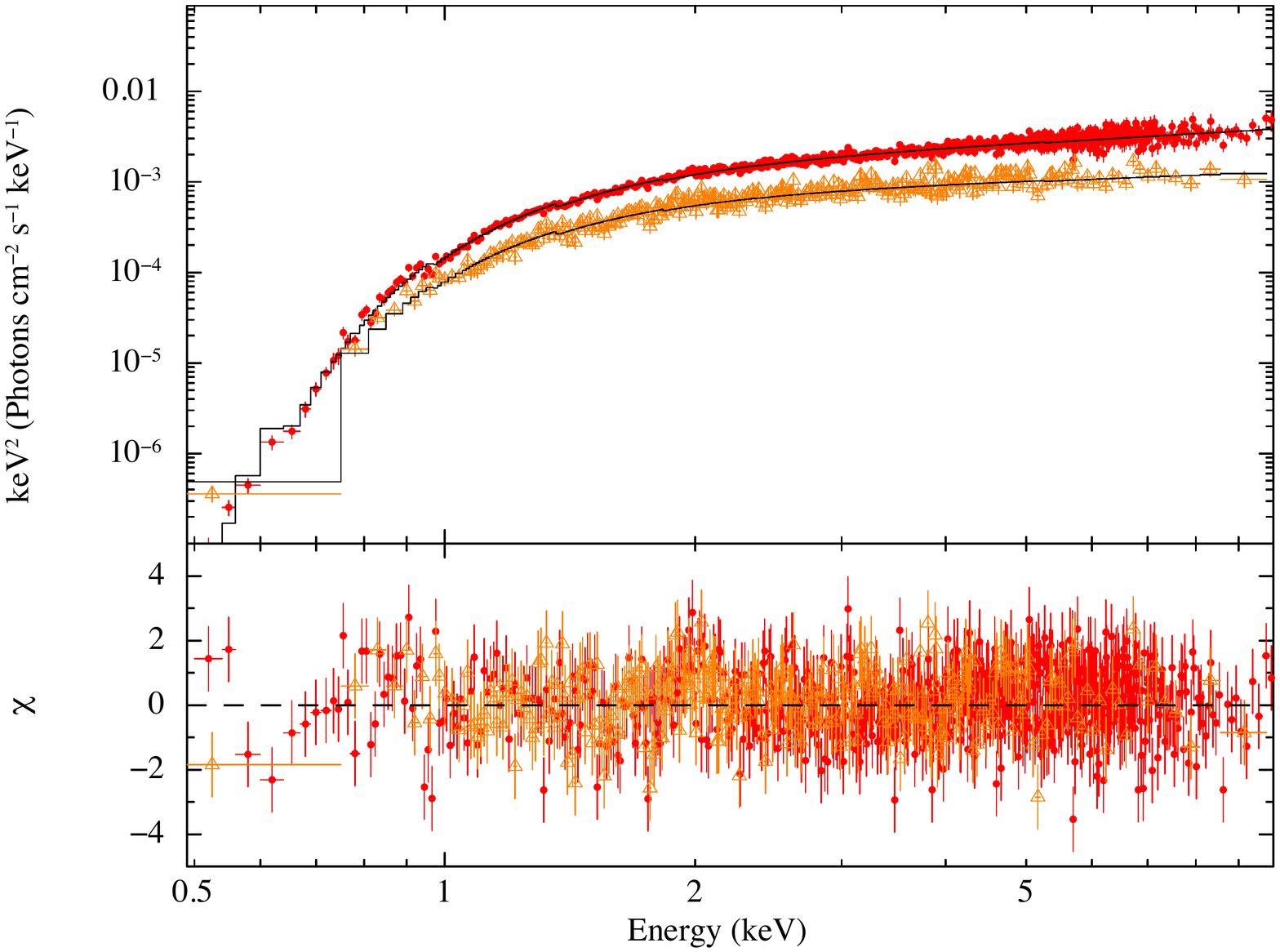} }
\caption[]{X-ray spectra 
obtained during the ``normal''decay, 
compared with the cumulative spectra
of the dips. 
{\bf Top panel:} WT mode observation 
from $T_0$+14.5 days to $T_0$+39.5 days: 
``normal'' spectra 
in blue circles and ``dip'' spectra 
in cyan triangles.
{\bf Bottom panel:} PC mode observation 
from $T_0$+40.5 days to $T_0$+404.8 days:
``normal'' spectra in red circles and  ``dip'' spectra in orange triangles. 
The corresponding spectra 
are clearly similar in shape, 
i.e. fit by the same model. See fit results in (see Table~4).
\label{fig:spectra}} 
\end{figure}

With the GTIs of the dips in Table~3  
we have been able to extract cumulative ``dip'' and ``normal'' 
spectra for each observing period, and corresponding backgrounds.
We fit the resulting four spectra with both an absorbed power-law 
and an absorbed log-parabola spectral model, and selected the best-fit model 
via an F-test as already done with time resolved spectra 
in Section \ref{subsec:spectra}.
The final best-fit parameter values are listed in Table~4, with 
corresponding confidence contours plotted in Figure~\ref{fig:contours}.
The final spectral fits are shown in Figure~\ref{fig:spectra}.
In both periods, the ``dip'' and ``normal'' spectra 
have the same shape
(absorbed log-parabola for WT spectra in period  {\it (a)}, 
and absorbed power-law for PC spectra in period {\it (b)})
and the same level of absorption (i.e. consistent values of \NH).
The values of the photon indices though indicate 
that ``dip'' spectra are generally softer. 
This is in agreement with the harder-when-brighter correlation on
short timescales described in Section \ref{subsec:spectra}.  
Moreover, there is evidence for larger intrinsic \NH\ 
in PC (i.e. at later times) than in WT (only marginal for
``dip'' spectra), in agreement with the trends observed in the
top panel of Figure~\ref{fig:param} and 
in the right panel of Figure~\ref{fig:ParvsFlux}.


%

%

\section{Discussion}
\label{sec:discussion}

We have calculated the definitive X-ray 
light curve of \transient\ in the 0.3$-$10 keV band
as observed by \swift/XRT in three years of follow-up
and monitoring since the detection of the outburst.

The overall shape of the light curve consists of a set
of initial spiky flares followed by an episode of steep
decay underlying variability on shorter time scales 
from about 2 to 6 days after the BAT trigger 
(see Figure~\ref{fig:flux_lc}, bottom panel), 
then a fast rebrightening by about an order of magnitude,
and finally a long fading phase with a power-law like 
average decay abruptly ending on 2012 August 17, 507 days
after the trigger. Without this sudden shut-off the source would have remained 
fairly bright for about 8$-$10 years. 

Despite several claims in the literature that 
the X-ray decay follows a $t^{-5/3}$ power law 
\citep[e.g.,][]{Metzger12_SwiftJ1644,
Lei13_SwiftJ1644,Kawashima13_SwiftJ1644,
Kumar13_SwiftJ1644,Barniol13_SwiftJ1644,
Zauderer13_SwiftJ1644,Shen14_SwiftJ1644,
Tchekhovskoy14_SwiftJ1644},
we find that this is not the case. 
Our modeling of the flux light curve from 6.3 to 507 days 
after the trigger (performed in Section \ref{sec:lcfit})
gives us an estimate of the decay slope of  $1.48 \pm 0.03$
and shows that the data may be almost equivalently described
with a slightly steeper slope when the decay start time is
earlier than the BAT trigger by $\Delta t_{\rm offset}$ days,
provided that $\Delta t_{\rm offset}$ is not larger than a few.

\subsection{Model Interpretation}
\label{subsec:models}

\citet{Zauderer13_SwiftJ1644} argue that the rapid decline of the
X-ray flux at $t\simeq 500$ days heralded the turning off
of the relativistic jet in \transient.
They suggest that at that time, the accretion
rate within the jet feeding the BH dropped below the
Eddington rate, ${\dot M}_{\rm Edd} \simeq 6 \times 10^{-3}$~\msunyr,
assuming $M_{\rm BH} \simeq 10^{6.5}$~\msun\
and a conversion efficiency between rest-mass energy and accretion luminosity
of $\epsilon_{\rm acc} \simeq 1$.
Prior to this time, the 
ratio of accretion power to jet power was constant,
as evidenced by the constancy of the observed power law decay index.
During the time it was seen by XRT, the X-ray luminosity
of \transient\ varied over a dynamic range of about a factor
of $\sim$6000, from
$\sim$$2.2\times 10^{48}$ erg s$^{-1}$
to  $\sim$$3.7\times 10^{44}$ erg s$^{-1}$.
Thus if the switch-off luminosity corresponded to Eddington,
at peak luminosity the accretion rate
onto the BH would have been $\sim$6000 times Eddington.
  
  If the conjecture by \citet{Zauderer13_SwiftJ1644} is correct,
one may obtain a constraint on the combination
of efficiency parameters relating accretion rate to jet power.
The jet power can be expressed as
\begin{equation}
  L_J = \epsilon_J {\dot M}_{\rm fb} c^2,
\end{equation}
  and the observed X-ray luminosity (assuming all the jet power is in X-rays) is
\begin{equation}
  L_{\rm X, iso} = { L_J \over { {1\over 2} {\theta_J}^2 }},
\end{equation}
 where $\theta_J$ is the jet opening angle.
  Hence,
\begin{equation}
  L_{\rm X, iso} =  { \epsilon_J {\dot M}_{\rm fb} c^2
        \over {{1\over 2} {\theta_J}^2 }   }.
\end{equation}
Assuming that the jet turns off when the accretion rate drops below Eddington, 
which happens at ${\dot M}_{\rm fb} = 6\times 10^{-3}$~\msunyr\
and  $L_{\rm X, iso} = 3.7\times 10^{44}$ erg s$^{-1}$,
we obtain the constraint
\begin{equation}
  \epsilon_J {\theta_J}^{-2} = 5.4.
\end{equation}

According to theory, it is not expected  a priori for jet power to track accretion rate
   in the disk.  If the jet is fed via the
canonical Penrose-Blandford-Znajek process, the jet power would be
\begin{equation}
     L_J = 1.2\times 10^{46} \ {\rm erg} \ {\rm s}^{-1}
      \Phi^2_{*,30} M^{-2}_{\rm BH, 6} {\omega_H}^2 f(\omega_H),
\end{equation}
   where $\Phi_{*,30}$ is the magnetic flux
  threading the BH event horizon in units of $10^{30}$ Gauss cm$^{2}$,
 $M_{\rm BH, 6}$ is the BH mass in units of $10^6$~\msun,
   $\omega_H=a/(1+\sqrt{1-a^2})$ is the dimensionless angular frequency of the BH horizon,
   and $f(\omega)=1+0.35{\omega_H}^2 -0.58{\omega_H}^4$
      is a high spin correction  \citep{Tchekhovskoy10,Tchekhovskoy14_SwiftJ1644}.
    For a spin parameter $a=0.9$ this becomes
\begin{equation}
  L_J = 5\times 10^{44} \ {\rm erg} \ {\rm s}^{-1}
         \Phi^2_{*,30} M^{-2}_{\rm BH, 6}.
\end{equation}
\citet{Tchekhovskoy14_SwiftJ1644} argue that
   in order for the jet power to track the accretion rate,
there must exist a ``magnetically-arrested'' accretion disk (MAD),
 in which the magnetic flux threading the BH
  is determined by the ram pressure of the accretion flow.
    For such disks  $\epsilon_J \simeq 1.3a^2$,
  where $a$ is the dimensionless spin of the BH.
   For a nominal jet opening angle $\theta_J \simeq 0.1$,
  our constraint $\epsilon_J {\theta_J}^{-2}=5.4$
  implies a spin
 $a\simeq 0.2$ if \citet{Tchekhovskoy14_SwiftJ1644} are correct.
  Also,  the MAD state may be consistent with super-Eddington accretion.

  At some point during the fallback of debris
   from the TDE  an accretion disk is expected
to form and to dominate the decay law,  due to its slower inherent
time scale enforced by an ever-expanding outer edge
  to the accretion disk \citep{Cannizzo90,Cannizzo11_SwiftJ1644}.
    \citet{Cannizzo11_SwiftJ1644} argued that  for \transient\
this transition occurred at $t\simeq 10$ days, such that for $t>10$ days 
the decay law would be flatter, $s\simeq \nicefrac{4}{3}$.
In this work we show that
the best fit to the decay of \transient\ appears to be
in line with $s = 1.5$. Moreover, the quality of the fit is
good; therefore $s= \nicefrac{5}{3}$ appears to be excluded, as does
the value $s= \nicefrac{4}{3}$ expected from an advective disk.  
As a caveat on drawing any strong conclusions based on our putative
slope $s = 1.5$ for the unabsorbed flux light curve, we note that we
are averaging over large amplitude variations in flux versus
time. Although we utilize a multi-time step technique and average
values in $\log F_x - \log t$ space, it may be that our 
measured $s$ value is not robust enough to warrant a detailed
comparison with either of the two canonical theoretical s values.
Our small inferred value for the fallback time  $t_{\rm fb}\la 1$ day
 appears to rule out the large $\Delta t_{\rm offset}$ values
  that would be required to substantially increase $s$.
 The theoretical decay law  $s= \nicefrac{4}{3}$
  \citep{Kumar08,Metzger08,Metzger09,Cannizzo09}
   is dependent on a super-Eddington disk and differs from the  $s= 19/16$
  expected from a standard thin disk \citep{Cannizzo90}.
 This flatter decay ($s\simeq 1.19$)
  can also be strongly excluded from our fit, and would not be expected anyway
   since ${\dot M} \gg {\dot M}_{\rm Edd}$ for \transient.

Two schools of thought have emerged concerning
the fallback time $t_{\rm fb}$ for \transient.
\citet{Cannizzo11_SwiftJ1644} and \citet{Gao12_SwiftJ1644} argue for  $t_{\rm fb} \la 1$  day,
which leads to the inference of a tidal disruption radius to periastron radius ratio
$R_T/R_P \simeq 10$
for the TDE, whereas \citet{Tchekhovskoy14_SwiftJ1644} and
\citet{Shen14_SwiftJ1644} find $R_T/R_P \simeq 1$.
\citet{Shen14_SwiftJ1644} also favor a BH mass  $M_{\rm BH} \simeq 10^4-10^5$~\msun, lower than most other groups.
They advocate  $t_{\rm fb} \ga 10$ days  and assume
that the   $s= \nicefrac{5}{3}$ decay starts at $\sim$10 days, which is not consistent with our results.
It is also important to note that \citet{Tchekhovskoy14_SwiftJ1644}
and \citet{Shen14_SwiftJ1644} simply use the \transient\ data
taken from the XRT website \citep{Evans09} which assumes a single ECF 
($4.8 \times 10^{-11}$~erg~cm$^{-1}$~count$^{-1}$) and reports the {\it observed} flux,
whereas we have used the time-dependent ECF for the unabsorbed flux, 
which is related to the intrinsic luminosity.  As we show in Figure~\ref{fig:ECF}, 
the unabsorbed flux ECF is about $9.6 \times 10^{-11}$~erg~cm$^{-1}$~count$^{-1}$, 
resulting in fluxes a factor of 2 higher than those obtained with the standard analysis; 
this affects the energetics, but not the derived decay slope $s$ of the light curve.

An important caveat to the results of \citet{Cannizzo11_SwiftJ1644} and \citet{Gao12_SwiftJ1644} has emerged
in the past few years, namely, these studies adopted a theoretical value for
  $t_{\rm fb}$  from \citet{Lacy82} and \citet{Rees88} that derives from
   relating the spread in specific orbital energy of the debris streams
to conditions at periastron, $\Delta\epsilon\simeq G M_{\rm BH}R_* {R_P}^{-2}$.
  However, recent work has shown that this viewpoint is not correct:
    the stellar shredding by the tidal force is in fact so effective
that by the time the star arrives at periastron, its shredded fragments
   are traveling on ballistic trajectories; therefore the relevant radius
  in determining  $\Delta\epsilon$ and thus  $t_{\rm fb}$ is not the
  periastron radius $R_P$,  but rather the disruption radius  $R_T$
  \citep{Guillochon13,Stone13}.
         This leads to smaller $\Delta\epsilon$ and larger $t_{\rm fb}$.
         However, the more recent studies do not consider strong general relativistic  effects  under the Kerr metric,
    which would become  relevant if an orbit
   were deeply plunging such that $R_P$ were to lie within the ergosphere
   of a nearly maximal spin  BH.  It is  conceivable  $t_{\rm fb}$   could be shortened dramatically.
Support for this idea may be given
by a recent study by \citet{Evans15} 
which presents a new class of TDEs showing prompt
formation of an accretion torus and hyperaccretion.
These TDEs involve ultra-close encounters ($R_T / R_P \simeq 10$)
and high spin BHs. They find a strong influence of general relativistic effects. 
A caveat to their work is that their large $\Delta \epsilon$ values 
may be an artifact of under-resolving the midplane compression of the star.

  In any event, the observational inference on the fallback time we derive
  in this work,  $t_{\rm fb} \la 1$ days,
     is based on two observables, the peak X-ray flux and the total X-ray fluence,
   and is therefore not subject to the theoretical uncertainties
 inherent in the aforementioned works.
  \citet{Krimm2011} note apparent activity in \transient\ on 2011 March 25,
 $\sim3$ days before the 2011 March 28 {\it Swift}/BAT trigger.   This interval of time exceeds
our nominal  $t_{\rm fb}$ estimate.
  The signal-to-noise (S/N for the  2011 March 25 was low,
   $\sim3.7\sigma$ ($0.0059 \pm 0.0016$ ct s$^{-1}$ cm$^{-2}$),
 but the positional coincidence with the larger
     trigger three days later,  S/N$=7.6\sigma$,  gives strength to the detection.
  A prior close encounter of the star that became disrupted
    may have led to a  partial disruption, such that an extended train
  of debris arrived near the BH prior to the main TDE, leading to a weak precursor.

\subsection{Residual X-ray Emission and Future Evolution}
\label{subsec:residual}

\transient\ is still detected 
at the flux level of (1.0$\pm$ 0.8)$\times$10$^{-14}$ erg~cm$^{-2}$~s$^{-1}$  
(0.3$-$10 keV, observed flux) over an integrated exposure
of $\sim$200 ks accumulated in $\sim$500 days of
weekly XRT monitoring.
The \chandra-ACIS ToO observation 
performed on 2012 November 26, $\sim$3 months after the XRT drop,
also detected \transient\ with 2.8$\sigma$ significance. A detailed
analysis of the observation is reported in \citet{Zauderer13_SwiftJ1644}.
We reproduced their results and estimated a 0.3$-$10 keV  unabsorbed flux\footnote{Flux error for \chandra\ points is at the 68\% confidence level.}  
of (7$\pm$3)$\times$10$^{-15}$ erg~cm$^{-2}$~s$^{-1}$,
consistent with the final \swift/XRT data point (see Figure~\ref{fig:flux_lc}). 
This flux value is obtained with the average late XRT spectral   
parameters estimated and used by \citep{Zauderer13_SwiftJ1644}: 
$\rm{N}_{\rm{Hgal}}$ fixed to 1.7$\times 10^{20}$  cm$^{-2}$,
intrinsic \NH$\sim 1.4 \times 10^{22}$ cm$^{-2}$, and $\Gamma \sim 1.3$.
A fit with \NH\ fixed to 1.9$\times 10^{22}$ cm$^{-2}$ gives $\Gamma = 2.1^{+1.7}_{-1.4}$
(consistent with the spectrum of our XRT last point presented in Section \ref{subsec:spectra})
and an unabsorbed 0.3$-$10 keV flux of  (8$\pm$3)$\times$10$^{-15}$ erg~cm$^{-2}$~s$^{-1}$. 
In the second observation performed by \chandra\ on 2015 February 17 
(day 1421 after the BAT trigger),  
\transient\ is still detected with 5$\sigma$ significance and net count rate 
of (4.2$\pm$1.2)$\times$10$^{-4}$ counts s$^{-1}$ in the 0.5$-$8 keV range. 
We modeled the spectrum with an absorbed power-law model as we already did with previous data. 
Fixing $\rm{N}_{\rm{Hgal}}$ to 1.7 $\times 10^{20}$  cm$^{-2}$, 
and intrinsic \NH\ to $\sim 1.4 \times 10^{22}$ cm$^{-2}$ 
we obtain $\Gamma = 0.6 \pm 1.2$, $\chi^2_r = 0.906$ (10 dof), 
while \NH\ fixed to 1.9$\times 10^{22}$ cm$^{-2}$  gives  
$\Gamma = 0.74\pm 1.2$, $\chi^2_r = 0.932$ (10 dof).
The unabsorbed flux in the (0.3$-$10) keV band is equal to 
$(1.7^{+0.4}_{-1.0}) \times $10$^{-14}$ erg cm$^{-2}$ s$^{-1}$. 
The 2015 \chandra\ spectrum seems to be harder than both the \swift-post drop spectrum 
and the \chandra\ 2012 spectrum, though given the large uncertainties on the photon indices 
the statistical significance is marginal.
Moreover, 
this 2015 detection is still consistent with the final \swift/XRT data point (see Figure~\ref{fig:flux_lc}).

The residual emission detected by \chandra\ and \swift\ 
is not consistent with thermal emission from the fall-back accretion disk.
For M$_{\rm BH}$ in the 5$\times$10$^6$$-$10$^7$~\msun range,
the disk is expected to have a temperature at the inner radius 
in the $\sim$20$-$25 eV range 
if jet shut-off occurred at a critical accretion rate 
$\sim$$\dot{\rm M}_{\rm Edd}$,
or in the $\sim$12$-$15 eV range 
if the critical accretion rate was $\sim$0.1~$\dot{\rm M}_{\rm Edd}$.
By using the {\tt diskbb} model in {\tt xspec} 
with true inner radius $\sim$6~G~M$_{\rm BH}$/c$^2$,
face-on disk, and redshift effects properly taken in account, 
we predict a 0.3$-$10 keV flux always $<$ 10$^{-16}$ erg~cm$^{-2}$~s$^{-1}$,
and a bolometric disk luminosity L$_{\rm disk}$ in the 
5$\times$10$^{43}$$-$10$^{45}$ erg s$^{-1}$ range.
By using the standard spectral model for Comptonized 
X-ray emission from an AGN corona described in \cite{Ghisellini09}, 
which consists of a power-law with photon index $\sim -$1,  
a high energy exponential cut-off with folding energy 
of $\sim$150 keV, and total X-ray luminosity $\sim$0.3~L$_{\rm disk}$, 
we can easily calculate that the observed 0.3$-$10 keV flux
could be obtained as a result of 
Comptonization of the disk UV photons by a hot corona
for an L$_{\rm disk} \simlt$10$^{44}$ erg s$^{-1}$. 
However, fine tuning of a number of  unmeasurable parameters
is required to achieve this result, and this makes the interpretation
of residual X-ray emission from \transient\ as Comptonized emission 
unlikely, though we cannot rule out contributions from Comptonization 
effects.

An alternate  interpretation is that 
the residual X-ray emission detected by \chandra\ and 
\swift\ after jet shut-off originates from the forward shock 
of the jet still expanding in the ambient medium. 
In fact, \citet{Zauderer13_SwiftJ1644} calculate  
that the low X-ray flux measured by \chandra\ 
is consistent with synchrotron emission from the forward shock 
of a structured jet with time-dependent physical parameters 
derived by modeling the radio 
spectra of \transient,
which they regularly monitored for about 600 days after the outburst.
In this scenario we expect the X-ray emission to go on
decaying with time as the radio emission. 
An extrapolation of the 2012 \chandra-ACIS flux to 2015
can be done assuming no spectral variation with time 
(as guaranteed by the extrapolated value of the cooling 
frequency based on \citet{Zauderer13_SwiftJ1644} results
still being in the NIR band),
and a flux decay rate $\propto t^{-\alpha}$ with $\alpha \sim (2-3p)/4$,  
where $p$ is the slope of the energy distribution of the electrons 
accelerated at the shock \citep{Granot02}. 
For a typical value of $p \sim 2.2-2.6$, we expect $\alpha$ in the $1.15 - 1.45$
range. Then, an unabsorbed flux lower than 
$\sim (2.6 \pm 1)$$\times$10$^{-15}$ erg~cm$^{-2}$~s$^{-1}$ 
in the 0.3 -- 10 keV band was expected in February 2015.
The 2015 \chandra\ detection clearly shows that the X-ray flux has 
not decayed according to this prediction.
Unfortunately, we have no  simultaneous information on radio 
emission level from \transient\ and we cannot tell if the two bands 
are really evolving in an independent way.
Only a coordinated X-ray and radio monitoring will be able to answer this question.

A distinctive signature of the two possible scenarios 
for the origin of residual X-ray emission,
i.e. disk/corona related Comptonized emission and 
forward shock related synchrotron emission,
may be the photon index of the X-ray spectrum $\Gamma$.
In the former case, $\Gamma$ is expected to lie in the $1.5 - 2.2$ range: 
values much lower than 1.5 correspond to nonphysical Compton $y$ parameters 
in standard inverse-Compton scattering scenarios for modeling the X-ray  
power-law emission from the corona \citep{Zdziarski_1990};  
values much larger than 2.2 are observationally unlikely based on 
detailed X-ray spectral analysis of the non-beamed AGNs \citep{Corral_2011,Vasudevan_2013}.
In the latter case, since the \citet{Zauderer13_SwiftJ1644} analysis shows that 
the cooling frequency of the shocked electrons was and is likely still expected 
to be in the infrared band, 
the spectral slope in X-rays should be $\Gamma =\nicefrac{p}{2} + 1$, and
for a typical value of $p \sim 2.2-2.6$, we expect $\Gamma \sim 2.1-2.3$ \citep{Granot02}.
Unfortunately, none of our \swift\ or \chandra\ X-ray spectra
has enough statistics to unambiguously constrain the photon index.
Even our indication in favor of a $\Gamma > 2$ given by the measure of the
XRT band ratio after the shut-off (see Section \ref{subsec:hr_lc}) 
does not allow us to decide between the two cases. 

\citet{Krolik11_SwiftJ1644} and \citet{Bloom11_SwiftJ1644} 
point out that the variability time-scales observed in \transient\ 
are orders of magnitude shorter than those found 
in the relativistic jets of blazars, which presumably 
have a similar mechanism \citep[Blandford-Znajek][]{BZ77}.
\citet{Tchekhovskoy14_SwiftJ1644} provide a model 
that seems able to account for all of the behavior seen 
in the X-ray light curve of \transient, including both 
the initial rapid and strong variability in 2011 March 
as well as the sudden shut-off in 2012 August.
The uncertainty regarding the long term evolution of the whole 
system after jet shut-off,
the nature and timescales of possible future disk transitions and
their observable signatures are related to the uncertainty in the physics
of disk accretion and to the detailed accretion disk models adopted 
by different authors. 
According to \cite{Tchekhovskoy14_SwiftJ1644} 
a reactivation of the jet may occur due to 
a further transition of the disk to the 
Advection Dominated Accretion Flow (ADAF) stage  
expected for $\dot {\rm M}$ $\simlt$~0.01~$\dot{\rm M}_{\rm Edd}$.
This phenomenon is expected in analogy  
to state transitions from high/soft to low/hard emission 
observed in Galactic microquasars, 
and observationally associated  with a jet revival \citep{Fender04}.
The jet reactivation, predicted to occur sometime between 2016 and 2022 
at an X-ray flux level of $\sim$$10^{-14}-10^{-13}$ erg cm$^{-2}$ s$^{-1}$
(depending on M${_{\rm BH}}$ and the type and the disrupted fraction 
of the passing star), is expected to be observable for months.
The new X-ray bright stage of \transient\ will likely be 
detectable by \chandra\ through all the reactivation period, 
and possibly even by \swift.
In the final analysis, such questions will be decided by observations, 
and to this end \swift\ continues to observe \transient\ regularly, 
watching for another flare-up of this fascinating object.

\subsection{Spectral Analysis and Dips}
\label{subsec:grbcompare}

We have shown that the shape and the variability timescales 
in the X-ray light curve of \transient\ are unlike other known
sources with relativistic jets along the line of sight.
The spectral behavior of \transient\ is peculiar as well.
For the first several days the source seems to follow a pure harder-when-brighter
correlation similar to that observed in blazars, but the later time
analysis we have done can be better explained in terms of a
superposition of a slow hardening of the spectra along
with the global light curve decay, and a harder-when-brighter correlation
tracking variability at shorter timescales.
The typical harder-when-brighter behavior in blazar flares can
be explained in terms of the blazar spectral model \citep{Celotti08}
by a rise of the External Compton (EC) component with the accretion rate,
associated with an independent different evolution 
of the Synchrotron Self Compton (SSC) component. 
It has also been observed that the slope of the correlation between the photon index and the 2$-$10~keV 
flux can be different from flare to flare, and can even disappear 
\citep{Vercellone11}. This may happen when the SCC and EC components 
increase proportionally with the accretion rate but maintain balance, 
so that luminosity rises achromatically.

\citetalias{Burrows11_SwiftJ1644} 
successfully explained the early broadband spectra of \transient\
with a blazar-like spectral model, but no EC component was required by our data.
The stringent VERITAS and {\it Fermi} upper limits on gamma-ray emission even 
required a suppression of the Self Compton peak of the spectrum through pair 
production.
The EC component in blazars can be produced by seed photons 
from the Broad Line Region (BLR) and/or seed photons from the accretion disk  
interacting with the relativistic electrons in the jet.
Actually, it is very unlikely that a BLR had time enough to form
in the case of \transient, but it is still possible that EC from disk
photons contributes to the soft X-ray emission of \transient,
though the component peak, expected at very high energies, 
must be highly suppressed.
On the other hand, \citetalias{Burrows11_SwiftJ1644}  (e.g. Supplementary Figure~15) 
showed that both the XRT spectrum extracted at the light curve minimum 
4.5 days post-trigger
and the later intermediate level spectrum extracted at 8 days
post-trigger have an upward kink at higher energies that suggest
the presence of an unknown additional hard spectral component, 
the peak of which they could not constrain.
This is also the case for all our XRT spectra fit by a concave log-parabola model.
This unknown hard component may play a role similar to the low energy
tail of the EC in determining the spectral evolution of \transient\
in X-rays.

The softness of the average emission of the dips 
compared to the inter-dip normal emission, 
without \NH\ variation, 
agrees with a production mechanism based 
on a larger contribution of a hard spectral component 
at larger accretion rate/luminosity.
The apparent randomness in 
temporal distribution, duration, and depths of the dips,
as well as in height, duration, and structure of the flares 
occurring between dips, suggest they originate from random
fluctuations in the accretion rate, i.e. instabilities in
the accretion flow, like in blazars.
However, dynamic range and duty cycle of dips and flares in 
\transient\ are extreme compared to blazars, 
and are difficult to reproduce in this model.
\citet{Krolik11_SwiftJ1644} suggest that the required
extremely compact and short-lived inhomogeneities (i.e. ``knots'') 
in the accretion flow,
can be obtained if the tidally disrupted star is a white dwarf
and the tidal disruption goes on taking away fragments of it 
at each periastron passage until it is consumed. However,  
only a central BH mass of $\sim$10$^4$ M$_{\odot}$ would fit the
observed timescales, which is not consistent with results
from our analysis of the global decay of the \transient\ 
light curve.
An alternative mechamism based on random internal shocks 
propagating along the jet channel has been proposed by 
\citet{DeColle12}.
Based on a hint of possible periodic modulation of the dips, 
\citet{Saxton12} proposed they may be due to combined effects 
of precession and nutation that cause the core of the jet 
briefly to go out of the line of sight.

\section{Conclusions}
\label{sec:conclusions}

   We present the definitive \swift/XRT
 light curve for \transient, which spans $\sim$800 days.
  We find that the ECF varies over time due to spectral
 evolution. Most
            previous studies of the long term  \transient\
          relied on a single ECF for the entire light curve.
   The peak flux, at $(t-T_0) = 1.3$ days,
  was $\simeq9\times 10^{-9}$ erg cm$^{-2}$ s$^{-1}$,  
    and the fluence over the  entire light curve
  was $\simeq6\times 10^{-4}$ erg cm$^{-2}$.  
   With the standard cosmology,
   a redshift $z=0.354$ yields peak luminosity and total energy values of
      $\sim2\times10^{48}$ erg s$^{-1}$
  and $\sim2\times10^{53}$ erg, respectively. 
       From the ratio of these numbers we determine an observationally based value for the fallback time for
debris following the TDE of $t_{\rm fb} \la 1$ days.
  By fitting the decay slope for 6 days $< t < $ 508 days 
  for six logarithmic binnings in $\log \Delta t$
  we determine a post-fluctuation decay slope
  $s=1.48\pm 0.03$.
   This is statistically distinguishable
   from the $s=\nicefrac{5}{3}$ value for \transient\
   commonly cited in the literature, and also from
   the  $s=\nicefrac{4}{3}$ value advocated by \citet{Cannizzo11_SwiftJ1644} 
   and \citet{Gao12_SwiftJ1644} 
   wherein one has a rapid transition
   from stellar fallback to highly advective disk accretion.
   Given the large fluctuations in X-ray flux with time,
   it may be difficult even with our multi-time step averaging
   technique to reliably extract a physically meaningful slope
   which bears comparison to theory.
   Previous studies quoting a slope
       did not carry out
           detailed fitting but
             simply overlay
            a $s=\nicefrac{5}{3}$ decay onto
  $\log f_X - \log t$ light curve for  \transient\ taken from the
    \swift/XRT
   website,  which assumes a single ECF value.

 Our small inferred $t_{\rm fb}$ supports the viewpoint of a rapid transition
     from stellar fallback to disk accretion 
    \citep{Cannizzo11_SwiftJ1644,Gao12_SwiftJ1644} 
     but the value of $s \approx 1.5$ does not.
     A value  $t_{\rm fb}\la $ 1 day  challenges current theory, which favors
  $t_{\rm fb} \simeq 20-30$ days, but does not  consider  strong general relativistic
      effects in the Kerr metric for large $R_T/R_P$ encounters;
         modifications in the binding energy spread $\Delta\epsilon$
          for the tidal debris
           from the standard results for $R_T/R_P \simeq 1$  encounters are
              treated  via linear perturbations to a Newtonian gravitational potential
              \citep[e.g.,][$-$ see their Section 6]{Guillochon13,Stone13}.

  For completeness we note that
  several recent works have addressed the issue of the fate of the
  shredded gas following the TDE \citep{Hayasaki13,Hayasaki15,Shiokawa15,Bonnerot15,Guillochon15}.
  \citet{Shiokawa15}
  consider the $R_T/R_P=1$  tidal disruption of a $0.64$~\msun\ WD by a
  $500$~\msun\  Schwarzschild BH utilizing a general relativistic
  hydrodynamic simulation (excluding magnetic fields).
  They find deflection of mass by shocks to be an important effect.
  The peak accretion rate is lowered by about a factor of ten compared
  to previous estimates and the duration of the peak is enhanced
  by about a factor of five. 
  \citet{Bonnerot15}
  investigate $R_T/R_P=1$ and  $R_T/R_P=5$
  tidal disruptions of a $1$~\msun\ star by a $10^6$~\msun\ Schwarzschild BH
  using an SPH (smoothed particle hydrodynamics) code. They find that
  circulation of debris is driven by relativistic apsidal precession
  (which causes the leading part of the stream to collide with the
   trailing part that is still falling back toward the BH). They
   consider two cooling efficiencies, isothermal and adiabatic
   equations of state.   
   \citet{Guillochon15} carry out simplified
   Monte Carlo realizations of tidal disruption streams
   to determine their structure prior to circularization. They find
   that for SMBHs with $a\ga0.2$ the stream self-intersection happens
   after the most bound debris has wound around periapse many times,
   and thus one expects the accretion rate onto the BH to be delayed
   with respect to the fallback rate, ${\dot M}_{\rm acc}(t+t_{\rm delay})
   = {\dot M}_{\rm fb}(t)$.  
   This delay would occur in addition to the one posited by 
   \citet{Tchekhovskoy14_SwiftJ1644}
   by considering magnetic flux build-up near the BH.  
   The results of \citet{Guillochon15} 
   are particularly interesting in light of our observational result
   $t_{\rm fb} + \Delta t_{\rm offset} \la 1$ day and indicate that
   additional physical effects may be required to address the unique
   aspects of \transient.
         
We cannot favor an interpretation of the late time low-level X-ray emission
   by \chandra\ and \swift\ as being consistent
   with emission from the forward shock of a structured jet because 
the second \chandra\ observation performed at the beginning of 2015 shows
that the X-ray flux has not decayed as expected in this scenario from
an extrapolation of the radio decay trend.
However, further coordinated X-ray and radio monitoring of the source
is needed to rule out the suggested common origin of the residual X-rays 
and the radio emission.
    The low-level X-ray emission is not consistent with thermal emission from 
the fallback accretion disk expected at these late times, 
but maybe reconciled with a scenario including Comptonized 
emission from a hot corona. 
   
   The spectral variability of \transient\ in X-rays can be described 
by an irregular harder-when-brighter behavior tracking flares and dips,
with a long term hardening trend associated with the decay phase.
The harder-when-brighter behavior may arise from the interplay
between the synchrotron spectral component and Comptonized radiation
from the accretion disk. 
Our study of the duration, depth and dynamic range of the dips 
in the time interval from $\sim$14 to $\sim$405 days post-trigger 
confirm extreme variability of \transient\ throughout all 
the decay phase of the light curve.

\acknowledgments

This work was supported by NASA grant NNX10AK40G. 
This work made use of data supplied by the UK Swift Science Data
Centre at the University of Leicester. 
We acknowledge the use of public data from the \swift\ data archive.

{\it Facility:} \facility{Swift}


\end{document}